\documentclass[]{aa}
\usepackage{graphicx}
\usepackage{txfonts}
\usepackage{natbib}
\begin{document}
\title{A deconvolution map-making method for experiments with circular scanning strategies}
\titlerunning{A deconvolution map-making method}
\subtitle{}
\author{D. L. Harrison\inst{1,3} \and F. van Leeuwen\inst{1} \and  M. A. J. Ashdown\inst{2,3}}
\institute{Institute of Astronomy, Madingley Road, Cambridge, CB3~0HA, UK \and 
Astrophysics Group, Cavendish Laboratory, Madingley Road, Cambridge, CB3~0HE,
UK \and Kavli Institute for Cosmology}
\date{Received: date / Revised version: date}

% \abstract{}{}{}{}{} 
% 5 {} token are mandatory

\abstract
{} 
{To investigate the performance of a deconvolution map-making algorithm for an experiment with a circular scanning strategy, specifically in this case for the analysis of {{\it Planck }} data, and to quantify the effects of making maps using simplified approximations to the true beams.}
{We present an implementation of a map-making algorithm which allows the combined treatment of temperature and polarisation data, and removal of instrumental effects, such as detector time constants and finite sampling intervals, as well as the deconvolution of arbitrarily complex beams from the maps. This method may be applied to any experiment with a circular scanning-strategy.}
{ Low-resolution experiments were used to demonstrate the ability of this method to remove the effects of arbitrary beams from the maps and to demonstrate the effects on the maps of ignoring beam asymmetries.  Additionally, results are presented of an analysis of a realistic full-scale simulated data-set for the {{\it Planck}} LFI~30~GHz channel.}
{Our method successfully removes the effects of the beams from the maps, and although it is computationally expensive, the analysis of the  {{\it Planck}}  LFI data should be feasible with this approach.}
 
\maketitle
%
%%%%%%%%%%%%%%%%%%%%%%%%%%%%%%%%%%%%%%%%%%%%%%%%

\section{Introduction}
\label{intro}

Recently, there has been a lot of activity in the development  of map-making methods for the European Space Agency satellite, {\it Planck} \citep{Planck_mission10, Planck_mission11}.  {\it Planck} has been designed to produce high-resolution temperature and polarisation maps of the cosmic microwave background (CMB). It has detectors divided between 9 frequency channels
sensitive to the frequency range of 30 to 857 GHz. These frequency channels are split between two instruments the HFI and LFI, the High and Low Frequency Instruments, respectively.  

Many of these map-making methods have been part of a coordinated development within the {\it Planck} collaboration, and have been tested using increasingly more sophisticated simulated data~\citep{ctp_mapmaking_one, trieste_paper,ctp_mapmaking_two}. The method presented here was developed with {\it Planck} in mind, but is applicable to any experiment with a circular scanning strategy.  It should be noted that this method is not, as of yet, part of the official data processing pipeline for the {\it Planck} project and has not been applied to the actual {\it Planck} data.

{\it Planck} spins about its axis once per minute, and as the line of sight
 (LOS) of the centre of the focal plane is almost perpendicular to this spin axis, the path of each detector
 describes an almost great circle on the sky.  The
 spin axis is repositioned at least once every hour, with the sequence of
 spin-axis positions defining the scanning strategy. The nominal path of the LOS for each rotation of the satellite reobserves the same almost-great circle on the sky, with the beams in the same orientations, for the duration of each pointing period.  A pointing period is the period of time between two sequential repositionings of the satellite spin-axis.  The nutation of the spin axis about its nominal position will produce variations in the LOS direction about the nominal path, changing the part of the sky which is observed. Provided the displacement of the LOS from its nominal path remains small with respect to the beam, the roughly 60 or so circles corresponding to a single pointing period may be thought of as a 1-dimensional ring on the sky and may be analysed together with our method. It should be noted that the LFI with its larger beams is inherently more robust to this issue, than the HFI. 

If the effects of beam asymmetries are not accounted for in the map-making
process, then this will result in systematic errors in the maps and in the
recovered power spectra. The systematic effects due to axisymmetric beams have
been assessed by \cite{carretti04} and a completely general assessment of the
asymmetries by \cite{odea07}. The map-making method described in this paper
provides a mechanism to account and correct for arbitrary beam asymmetries,
hence the term deconvolution map-making. Our approach is not the only one to
this problem; \cite{armitage04} have developed a method to account for the
effects of the asymmetries, which is less computationally expensive than our
method in the case of simple beam asymmetries, but is effectively restricted
in the detail of the beam description that can be implemented. Here we may
account for any degree of beam complexity. Both methods may also remove from the data
the instrumental effects due to detector time constants and the finite
sampling interval, whereby each data point is formed from the signal
integrated over a small time interval.

Removing the effects of the beam from the map could be  extremely useful for the study of resolved objects, which would otherwise be distorted by the beam. Additionally, removing the distorting effects of the beam could aid in studies of the lensing of the CMB \citep{lensing_review, lensing_planck}.

We have followed two approaches, to solving the map-making equations, in the development of our deconvolution map-making method: a direct approach from which the full noise covariance matrix may be recovered at low resolution and an iterative approach, which although still computationally expensive, will allow the resolutions required for the analysis of the LFI data to be reached. Both our approaches may be used to analyse polarised data with arbitrary beams, and recover the underlying multipoles on the sky without the need to pixelate the data.

We discuss the methods used and the implementation of our deconvolution map-making method in Section~\ref{method_imp}, from the preprocessing of the ring data in Section~\ref{tod2fc} to the reconstruction of the sky in Section~\ref{sphere_recon}. The simulations generated to fully test our method in the case of arbitrarily complex beams are described in Section~\ref{own_sim} together with the results of this analysis. The results of the analysis of  more realistic {\it Planck}  data, used in a previous map-making comparison paper \citep{trieste_paper},  is  described in Section~\ref{trieste_sim}.

\section{Method and Implementation}
\label{method_imp}

Our method splits the data processing in a way which is natural in terms of how {\it Planck} acquires the data, allowing the time-ordered data (TOD) from each pointing period to be processed separately. The TOD corresponding to each pointing period may be reduced, without loss of information, to Fourier coefficients on the rings as described below in Section~\ref{tod2fc}. These Fourier coefficients together with the mean pointing information for each ring may then be analysed to recover the multipoles on the sphere, as outlined in Section~\ref{sphere_recon}. 

\subsection{TOD to Fourier Coefficients on rings}
\label{tod2fc}
 
The time-ordered data for each pointing period is processed broadly as described in \cite{leeuwen02}. The implementation of this method to more realistic simulated data, however, has necessitated some modifications to the method presented in that paper. In this section we outline the processing required to extract the Fourier coefficients from the TOD, highlighting these modifications.

\begin{figure}
\begin{center}
\setlength{\unitlength}{1cm}
\begin{picture}(8,8)(0,0)
\put(-1.7,-1.7){\includegraphics{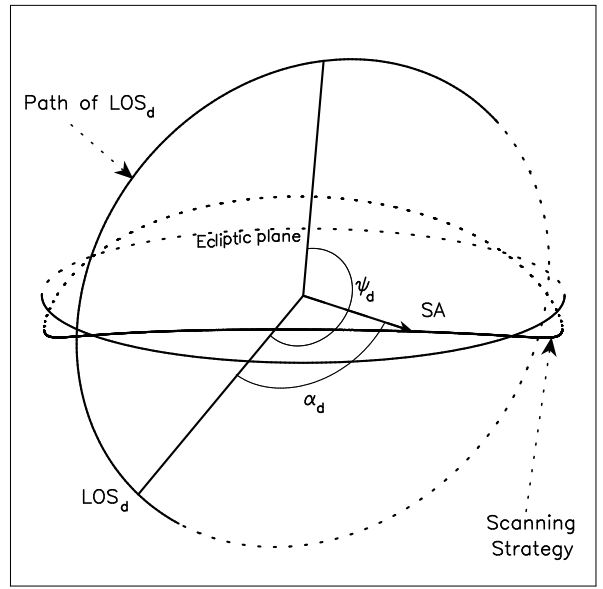}}
\end{picture}
\end{center}
\caption[]{Path described over a pointing period by the line-of-sight of a detector for a given opening angle, $\alpha_{d}$ and spin axis position. The position of the instantaneous line-of-sight may be described with the addition of another angle, $\psi_{d}$, the phase. The repositioning of the spin axis position over the course of the mission defines the scanning strategy; plotted here is the cycloidal scanning strategy which is used throughout this paper.} 
\label{tod_fig}
\end{figure}

The position of the line-of-sight of a detector for a pointing period may be described in terms of the mean spin axis position and two angles, as shown in Figure~\ref{tod_fig}. These angles are the opening angle, which is the angle between the spin axis position and the line-of-sight of the detector, and the phase, which defines the position around the ring from a given reference point. By assessing the value of the phase for each sample in the TOD, these data may be binned in phase, which effectively compresses the data by a factor of 40 to 50. If a suitable number of phase-bins is chosen, it is possible to recover the Fourier coefficients, which represent the TOD, from the phase-binned data with a negligible loss of accuracy as compared to evaluating them directly from the TOD, but with a significant reduction in the processing required.

It should be noted that our implementation differs from that in  \cite{leeuwen02} as to the number of phase-bins required, and the number of moments which are included in equations (4)-(8) in that paper. Our investigation showed that including the $\rm{3^{rd}}$ order terms of the phase improved the recovery of the Fourier coefficients and that there was no loss of accuracy in reducing the number of phase-bins from $6n_{max}$ as used in \cite{leeuwen02} to $4n_{max}$, where $n_{max} $ is the highest mode extracted.  The value of $n_{max}$ chosen should be such that $n_{max} \ge \ell_{max}$, where $\ell_{max}$ is the desired value to which the sky multipoles are to be recovered.

The above procedure successfully recovers the required Fourier coefficients
with negligible loss of accuracy when the distribution of the TOD is
relatively uniform in phase. If there is a resonance between the spin and
sampling rates then the distribution of the data in phase will no longer be
uniform, as samples from a previous rotation will occur at the same location
in phase as the current rotation. 

The performance of this method at recovering the Fourier coefficients from data with a non-uniform distribution in phase has been investigated. The degree of non-uniformity in phase,  in terms of the maximum gap present between two subsequent phase-ordered samples, which can be tolerated before the effect on the recovered values for the Fourier coefficients becomes significant with respect to the noise on the data, was assessed. Should this phase-gap size be exceeded, the Fourier coefficients will be evaluated directly. In the case of the simulations, described in Section~\ref{trieste_sim}, this meant that 5\% of the rings were not phase-binned, with the Fourier coefficients being evaluated directly from the TOD for these rings.

Although, when possible we phase-bin the TOD to reduce the processing requirement, the effect of this binning is negligible, as the Fourier  coefficients recovered from the phase-binned rings are of a negligible difference from those evaluated directly from the TOD. Therefore there can be no effects in our data analysis due to the binning or pixelisation of the data, which is one way in which our method differs from that of \cite{armitage04}.

\subsection{Sky reconstruction}
\label{sphere_recon}

The data in terms of the Fourier coefficients for a single ring, $d_r$, may be
expressed as
\begin{equation}
d_r=R_ra +n_r
\label{fc_eqn}
\end{equation}where $n_r$ represents the noise on the Fourier coefficients, $R_r$ is the coupling matrix which describes the connection between the multipole moments, $a_{\ell m}$ on the sphere, represented by the vector {\it a } and the Fourier coefficients.

The data from all the rings may be combined and analysed together
\begin{equation}
d=Ra +n
\end{equation} where $R$ is the equivalent of the pointing matrix in conventional pixel-based map-making. The maximum likelihood estimates for multipoles on the sky may then be found by solving the matrix equation

\begin{equation}
\left( R^T N^{-1} R \right) {\hat a} = R^T N^{-1} d
\label{matrix_eqn}
\end{equation} where {\it N} is the noise covariance matrix of the Fourier
coefficients, given by $\left< n n^T \right>$.

The coupling matrix is derived in \cite{challinor02} and requires information
on the detector orientations, opening angles and beam profiles, together with
the mean spin axis positions for each ring.  The coupling matrix is
constructed as in \cite{challinor02} with the exception of accounting for
those effects on the data which occur around the rings such as those due to sampling intervals and the detector time-constants, as removing these effects involves adjusting the Fourier coefficients for each pointing period, and is more naturally included in TOD to Fourier coefficient code, rather than in the sky reconstruction code.

As shown by \cite{challinor02} the correlations in the noise between different
Fourier modes on the rings are expected to be negligible. This expectation was
verified and we therefore treat the noise covariance matrix, {\it N}, for those data as diagonal.  This method could use the full noise covariance for each ring, making {\it N} block diagonal, however this is not necessary as gaps in the data, due to glitches or otherwise, should not occur in any preferential location, hence the number of samples per phase bin should be on average the same. This will ensure the near stationarity of the noise at the ring level, and negligible level of noise correlations between Fourier modes.  In the case of an experiment which does not have this redundancy, this method would still be applicable provided that a block diagonal {\it N} is used.

The presence of $1/f$ noise in the data results in striping in the maps, due to a  different offset on each ring \citep{burigana97,delabrouille98}.  If the knee frequency is low then by removing the zero-frequency Fourier coefficients from our analysis we may `destripe' the maps, effectively removing the contribution of the $1/f$ noise, and projecting out correlated noise on time-scales longer than a ring.  Given that the noise covariance matrix is diagonal, the zero-frequency Fourier coefficients may be removed from the analysis by setting the diagonal elements of the inverse noise covariance matrix which correspond to these coefficients to zero. This is equivalent to increasing the noise on these coefficients to infinity and hence their contribution to the recovered signal is completely removed. An alternative approach is to introduce an additional parameter for every ring and solve for the offsets produced by the $1/f$ noise, as well as recovering
the  signal multipoles. As one would expect, there is no change in the recovered values of the  signal multipoles using this approach, so this should only be used if the offsets themselves are required.  If the knee frequency is not sufficiently low to isolate the $1/f$ noise in the zero frequency modes, then the approach presented here can be extended to deal with any arbitrary noise power-spectrum on the rings, by suitably weighting the higher frequency modes.

The matrix, $\left( R^T N^{-1} R \right)$, the inversion of which is required
to obtain the solution for the spherical multipoles, unfortunately, is non-sparse and large with the order of $\ell_{max}^4$ elements. This effectively limits the maximum value of $\ell$ to which the analysis may proceed, as the computational requirements needed to produce a solution scale as ${\mathcal O} \left ( \ell_{max}^6 \right)$ for  a  direct method and ${\mathcal O} \left( \ell_{max}^4 \right)$ for iterative methods such as the preconditioned conjugate-gradient method which we have implemented.  It should be noted that there is an implicit assumption of full-sky coverage and a minimum of $4 \times \ell_{max}$ rings, should this not be the case the condition number of $\left( R^T N^{-1} R \right)$ will increase and in practice not all multipoles will be recoverable.

Since the coupling matrix is already non-sparse there are no increases in the
computational expense of our code if we choose to analyse the data with beams
whose expansion in spherical multipoles contain arbitrary high values of $m$,
in contrast to the method of \cite{armitage04}.

%%%%%%%%%%%%%%%%%%%%%%%%%%%%%%%%%%%%%%%%%%%%%%%%%%%%%%%%%%%%%%%%%%%%%%%%%%%
%% Figure with maps of the 3 different beams used
%%%%%%%%%%%%%%%%%%%%%%%%%%%%%%%%%%%%%%%%%%%%%%%%%%%%%%%%%%%%%%%%%%%%%%%%%%%
\begin{figure*}
\begin{center}
\setlength{\unitlength}{1cm}
\begin{picture}(16,18.5)(0,0)
\put(0,-0.5){\includegraphics{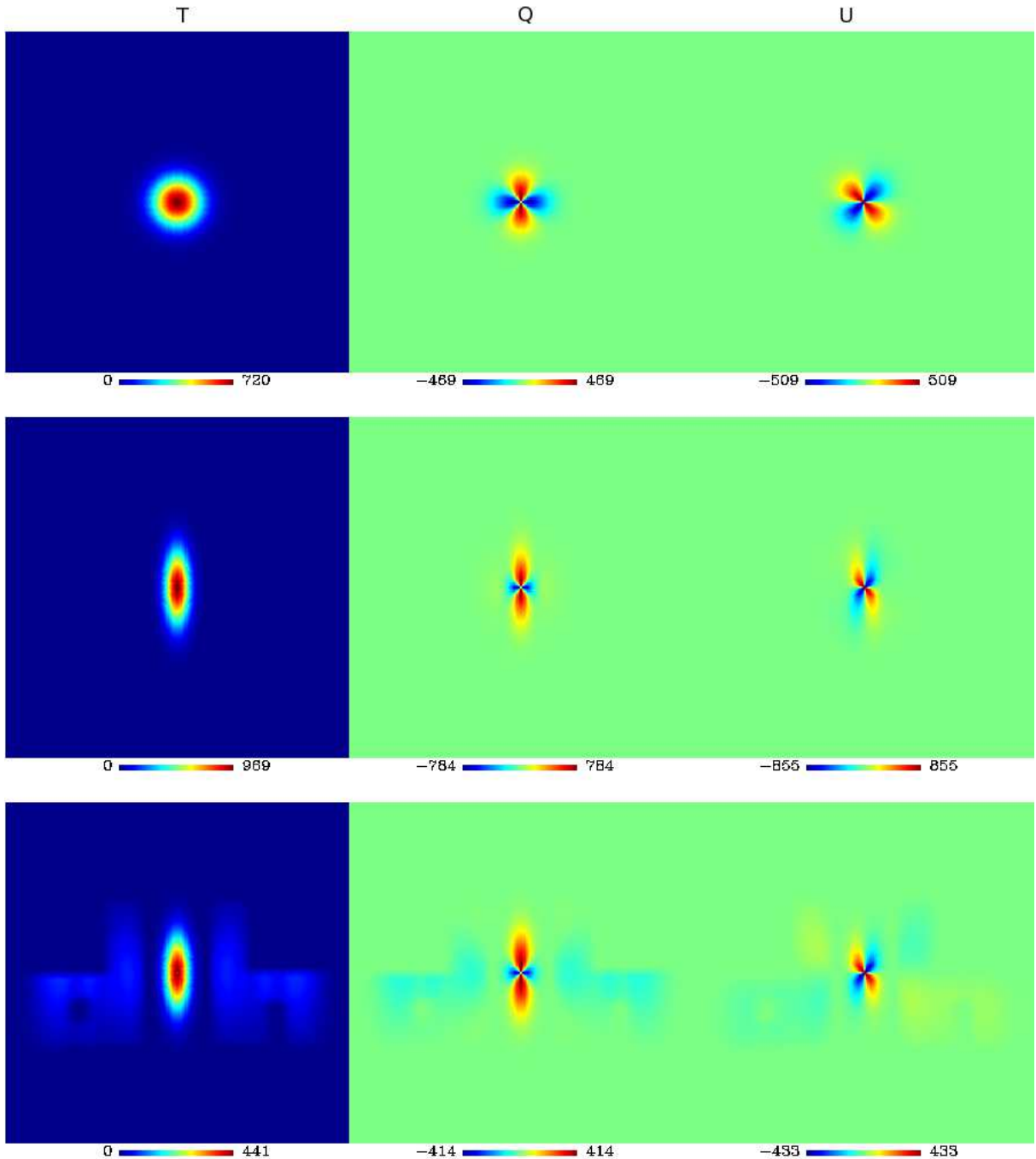}}
\end{picture}
\end{center}
\caption[]{$T$, $Q$ and $U$ maps, made using HEALPix\footnotemark\,  \citep{gorski05}, of the beams used in the complex-beam simulations. Top: the circularly symmetric beam; middle: the elliptic beam; bottom: the true beam for which the elliptic beam forms the main beam.}
\label{mybeams_fig}
\end{figure*}
%%%%%%%%%%%%%%%%%%%%%%%%%%%%%%%%%%%%%%%%%%%%%%%%%%%%%%%%%%%%%%%%%%%%%%%%

\subsubsection{Iterative Method}
\label{approx_section}

In order to reach the values of $\ell$ required for an analysis of {\it
  Planck} data, it was necessary to develop an iterative method for solving
equation~(\ref{matrix_eqn}) for which a parallel implementation is possible. Due to its large requirement on memory, the coupling matrix, {\it R}, must be stored over multiple processors, as a copy on each processor would be prohibitively expensive in terms of memory used; even stored once the size of {\it R} becomes prohibitive for large values of $\ell_{max}$ for the full mission analysis.

In order to reduce the memory requirements, {\it R} is evaluated as needed and only the part which corresponds to a single ring is stored in memory at any one time. This reduces the memory requirement to ${\mathcal O} \left( \ell_{max}^3 \right)$. The division of the storage and calculation of {\it R} between the processors should be such that it minimises the amount of data which must be exchanged between them. This requirement may be met by storing and evaluating {\it R} in terms of the sub-matrices corresponding to individual $\ell$ values. This division of the processing has implications for the scaling of the code with the number of processors used, which is described in Appendix~\ref{implementation_details}.

We use a preconditioner, the purpose of which is to achieve a reduction in the
number of iterations required at the expense of a small increase in the
computational cost of each iteration in terms of an additional matrix-vector
multiplication \citep{pcg_ref}. Our chosen preconditioner is the diagonal of
the matrix $ R^T N^{-1} R $, which meets these criteria.

The implementation of the preconditioned conjugate-gradient method described
here was parallelised using the Message-Passing Interface (MPI), and hence is
capable of being run on both shared-memory machines and clusters. 

\subsubsection{ Direct Method}
\label{direct_method}

Our direct method for solving equation~(\ref{matrix_eqn}) ,  takes advantage of the fact that in the case of a diagonal {\it N}, it is possible to pre-whiten the data, {\it d}, and the coupling matrix, {\it R}, so that equation~(\ref{matrix_eqn})  reduces to a standard least squares equation, which may be solved by QR decomposition and backsubstitution. The direct method performs the QR decomposition using Householder transformations \citep{newhip} to reduce {\it R} to an upper triangular form, {\it U}. This method allows the processing of {\it R} to be split into sections in terms of rows, provided that the number of rows of {\it R} being processed together is larger than the number of columns of {\it R}. Our implementation uses this property to ensure that the subsection of data being processed together with its corresponding section of the coupling matrix will fit within the available memory. As each subsection of data is processed the upper triangular matrix, {\it U}, is further refined and updated.

Once {\it U} has been found the $a_{\ell m}$ may be evaluated through backsubstitution. Additionally, $ U^{-1}$ may also be found through backsubstition, and hence the full noise covariance matrix for the  $a_{\ell m}$ may be evaluated. This covariance matrix may be useful as an input to the hybrid approach of \cite{hybrid_paper} for power spectrum or parameter estimation, which uses a direct likelihood evaluation at low multipoles and pseudo-$C_{\ell}$ estimators for high multipoles.

The direct approach was also parallelised using MPI, for reasons of code portability. Additionally, it makes use of the subroutines which calculate the sub-matrices of {\it R} for the individual values of $\ell$, and hence the direct method will be subject to the same scaling conditions on the performance with the number of processors used, described in Appendix~\ref{implementation_details}, as the iterative method.

\section{Complex-beam simulations}
\label{own_sim}

%%%%%%%%%%%%%%%%%%%%%%%%%%%%%%%%%%%%%%%%%%%%%%%%%%%%%%%%%%%%%%%%%%%%%%%%%%%
\footnotetext{http://healpix.jpl.nasa.gov}
%%%%%%%%%%%%%%%%%%%%%%%%%%%%%%%%%%%%%%%%%%%%%%%%%%%%%%%%%%%%%%%%%%%%%%%%%%%

As a validation step, in order to demonstrate the ability of our method to remove the effects of any arbitrarily-complex beam and to test it independently of the TOD to Fourier coefficient step, it was necessary to generate our own set of simulations.

The Fourier coefficients on the rings may be simulated directly by using equation~(\ref{fc_eqn}). An arbitrary beam was generated based on the first author's initials with the middle initial, {\it l}, being represented by an elliptical beam with major and minor full-width half-maximums (FWHMs) of 3 and 1 degrees, respectively. The largest sidelobe is -15.8 dB relative to the peak, and located 5.9 degrees from it. The set of Fourier coefficients on the rings, generated using this arbitrary beam, shall be referred to from now on as the complex-beam simulation, and this beam shall be referred to as the true beam.

The data for the complex-beam simulation consists of 800 rings, for two polarised detector pairs, containing Fourier coefficients up to an $\ell_{max}$ of $200$. These simulations are noise free; no noise is added to the coefficients, and a suitably low-level of noise is used to produce the noise covariance matrix, $N$.

To complete the set of beams with which to analyse the complex-beam simulation, the spherical multipoles for an elliptical and a circular beam were produced. The elliptical beam corresponds to the elliptical main lobe of the true beam, and a circular beam has a FWHM corresponding to the geometric mean of the two FWHMs of the elliptical beam. All the beams are defined by their spherical multipoles. However, for visualisation purposes, $T$, $Q$ and $U$ maps for these beams were produced. These maps, synthesised at the north pole of the coordinate system, may be seen in Figure~\ref{mybeams_fig}.

\subsection{Analysis of complex-beam simulations}
\label{in_house_sim_results}

\begin{figure}[ht]
\begin{center}
\setlength{\unitlength}{1cm}
\begin{picture}(8,22)(0,0)
%% input
\put(9,15){\includegraphics{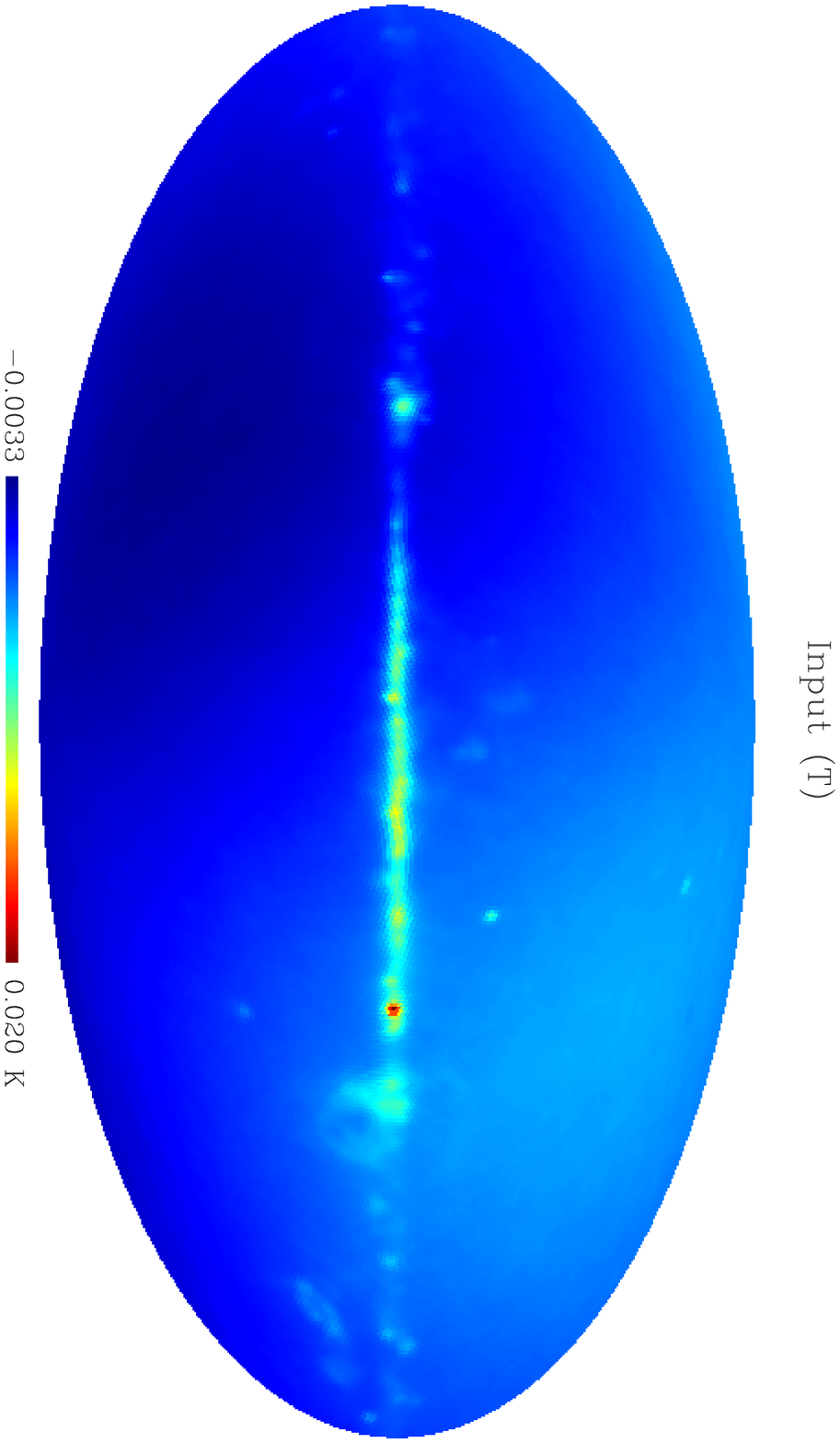}}
%% true
\put(9,9.5){\includegraphics{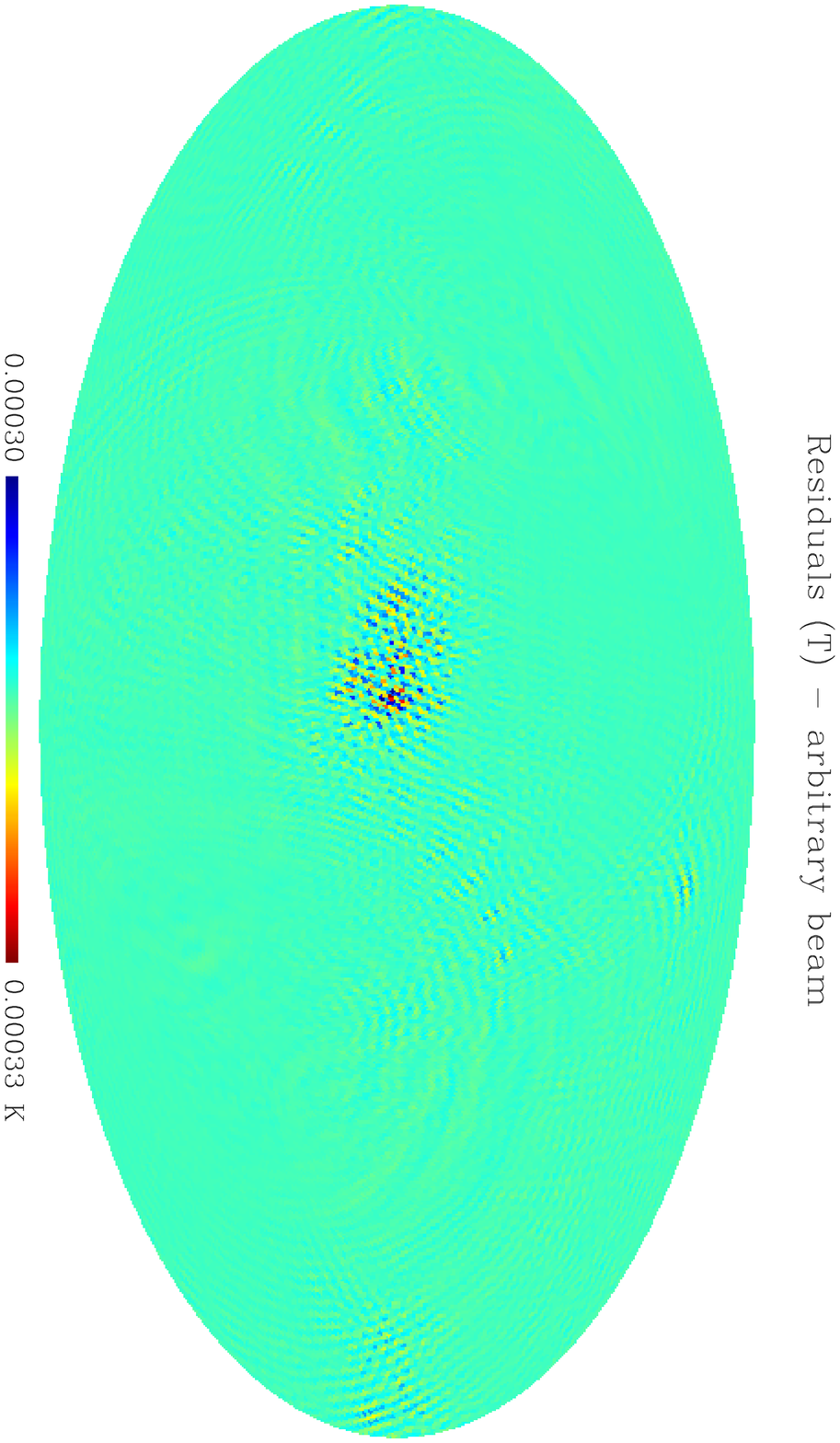}}
%% elliptical
\put(9,4){\includegraphics{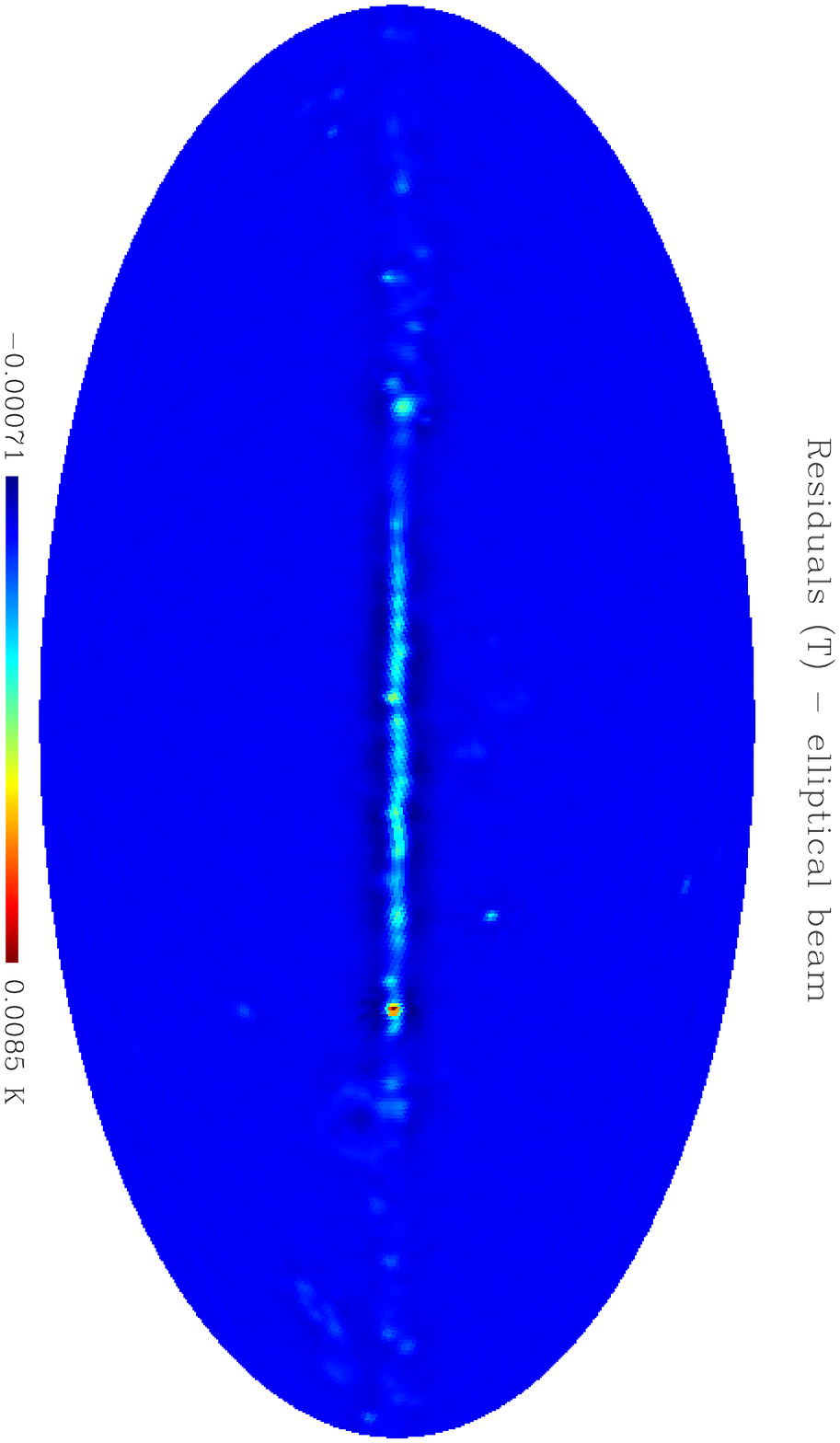}}
%% circular
\put(9,-1.5){\includegraphics{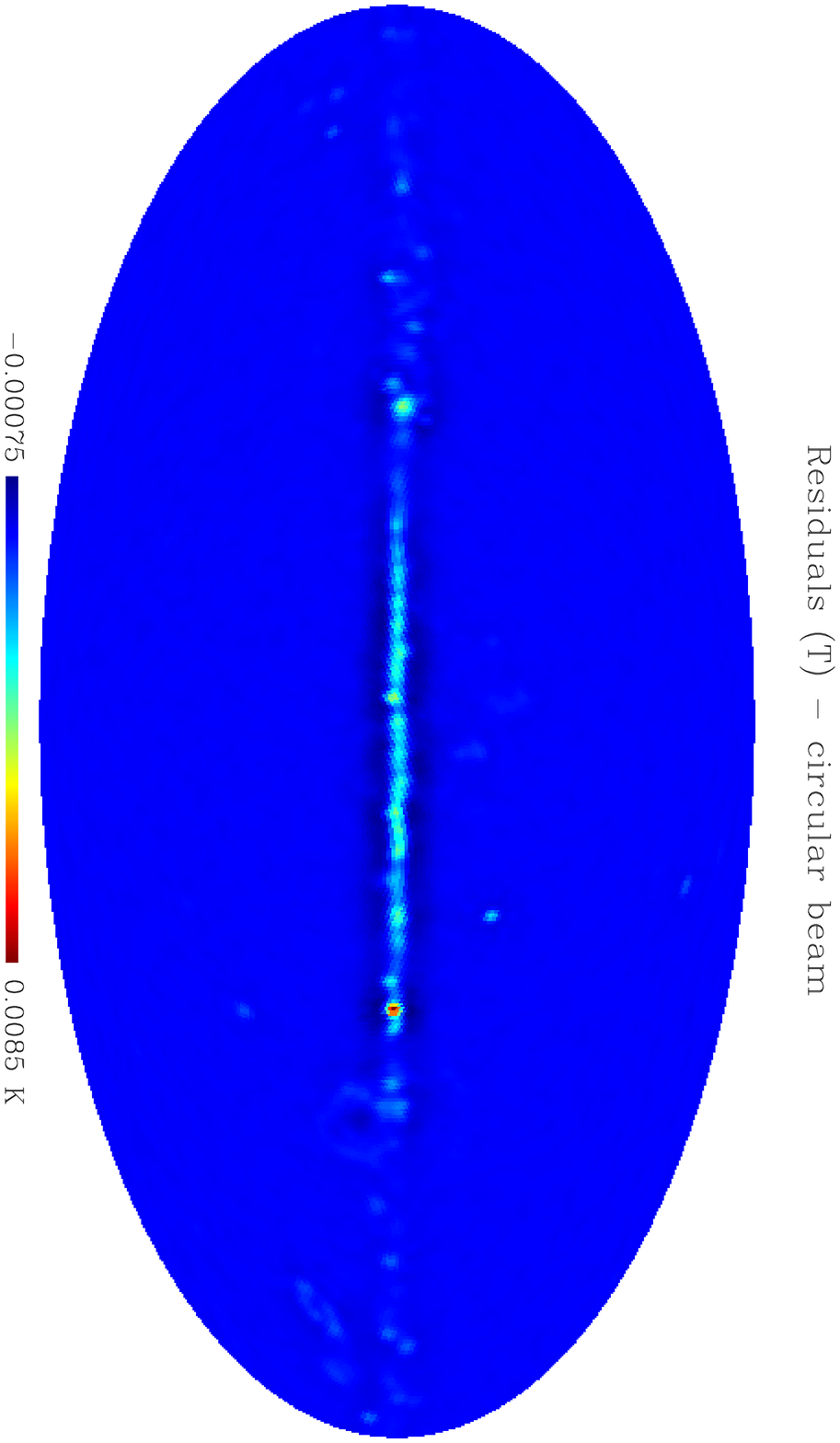}}
\end{picture}
\end{center}
\caption[]{Top panel shows the input temperature multipoles, $a^T_{\ell \it m}$  synthesised onto a HEALPix map of $\rm{ N_{side}}$=64. The lower panels show the residuals between the input and the recovered temperature multipoles, in the cases of the three different beams, from top to bottom the residuals maps are from the analysis with the true, elliptical and circular beams. Note that each residual map has a different colour scale.}
\label{inhouse_resT_fig}
\end{figure}

\begin{figure}[ht]
\begin{center}
\setlength{\unitlength}{1cm}
\begin{picture}(8,22)(0,0)
%% input
\put(9,15){\includegraphics{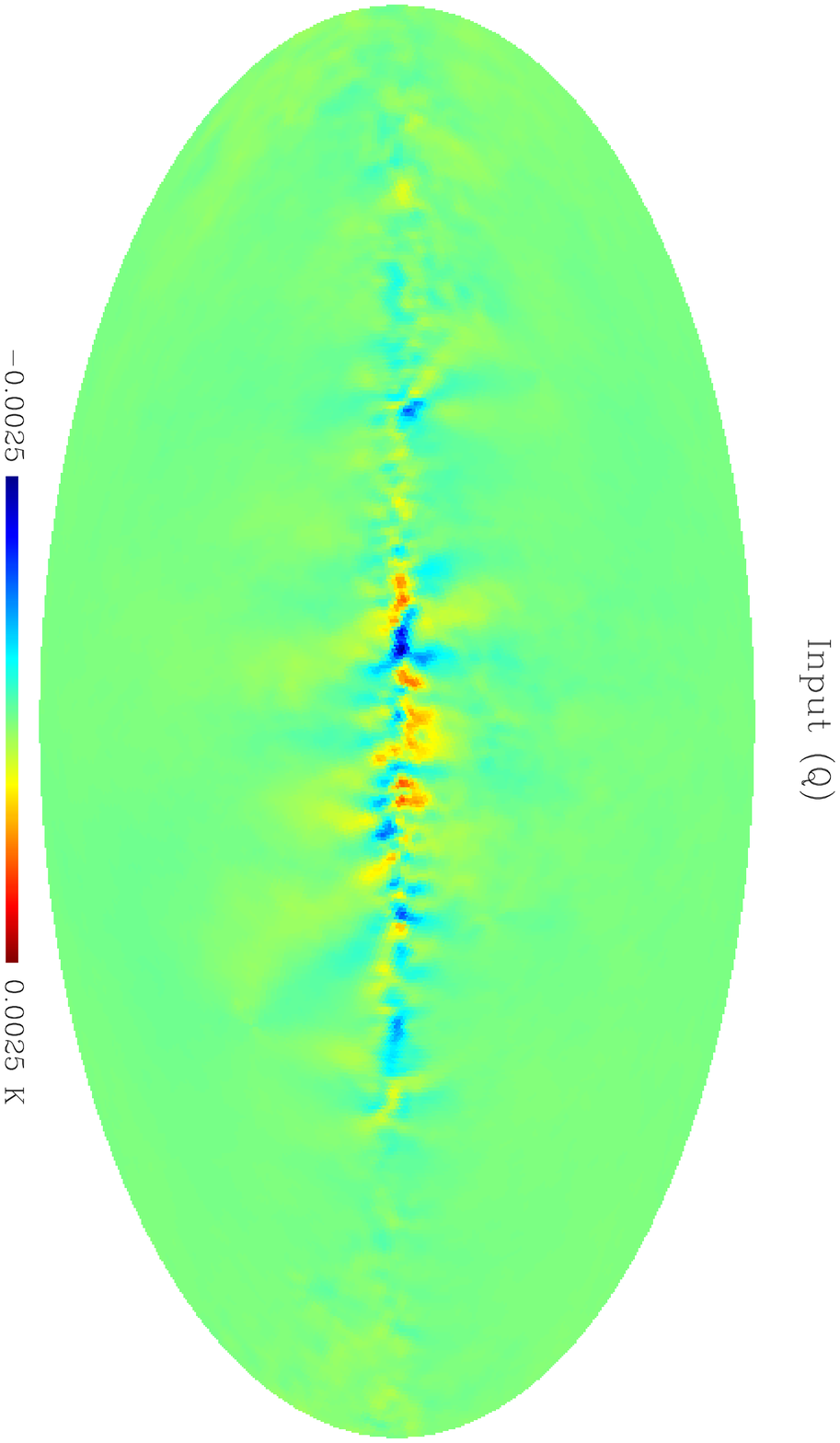}}
%% true
\put(9,9.5){\includegraphics{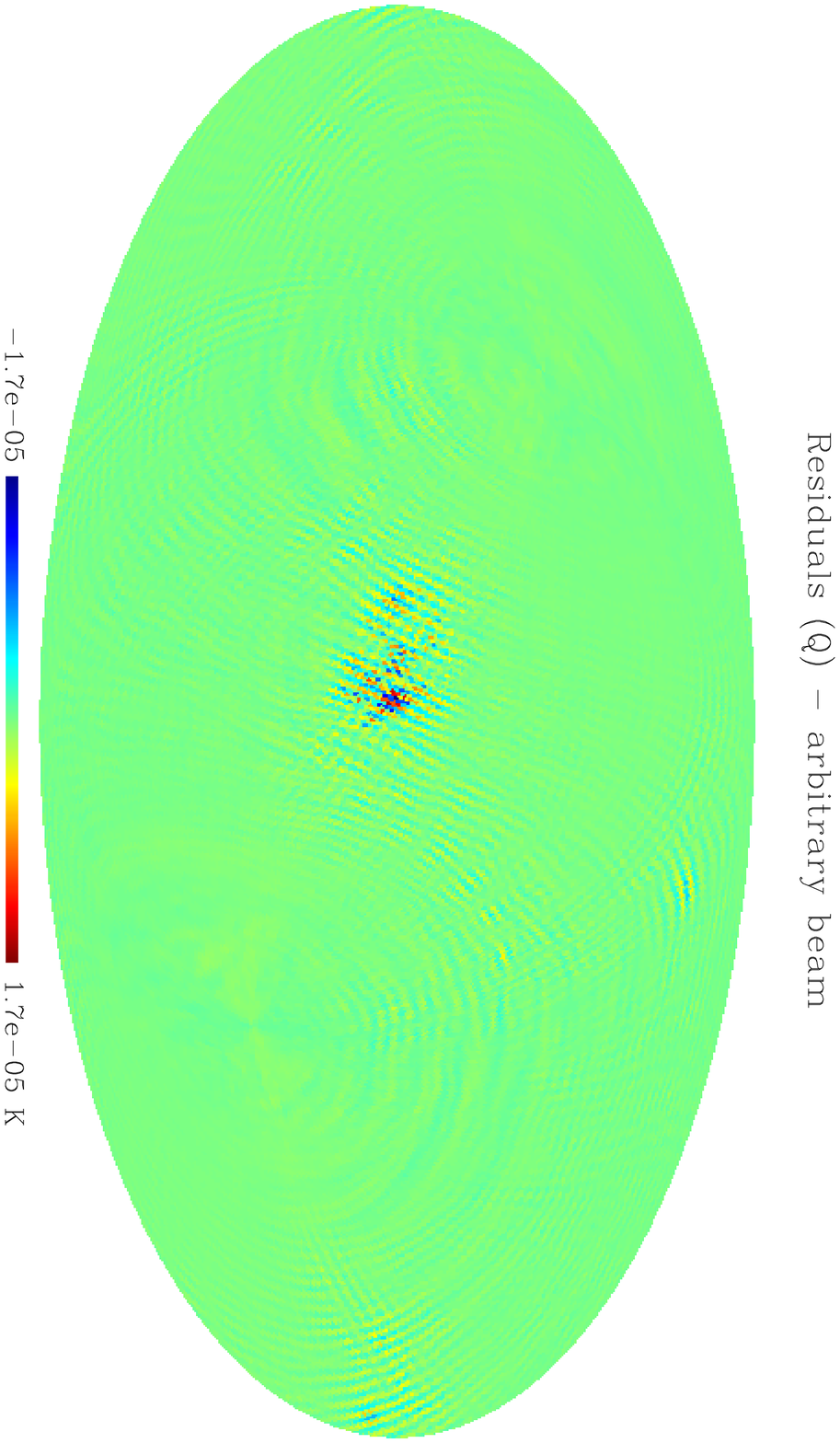}}
%% elliptical
\put(9,4){\includegraphics{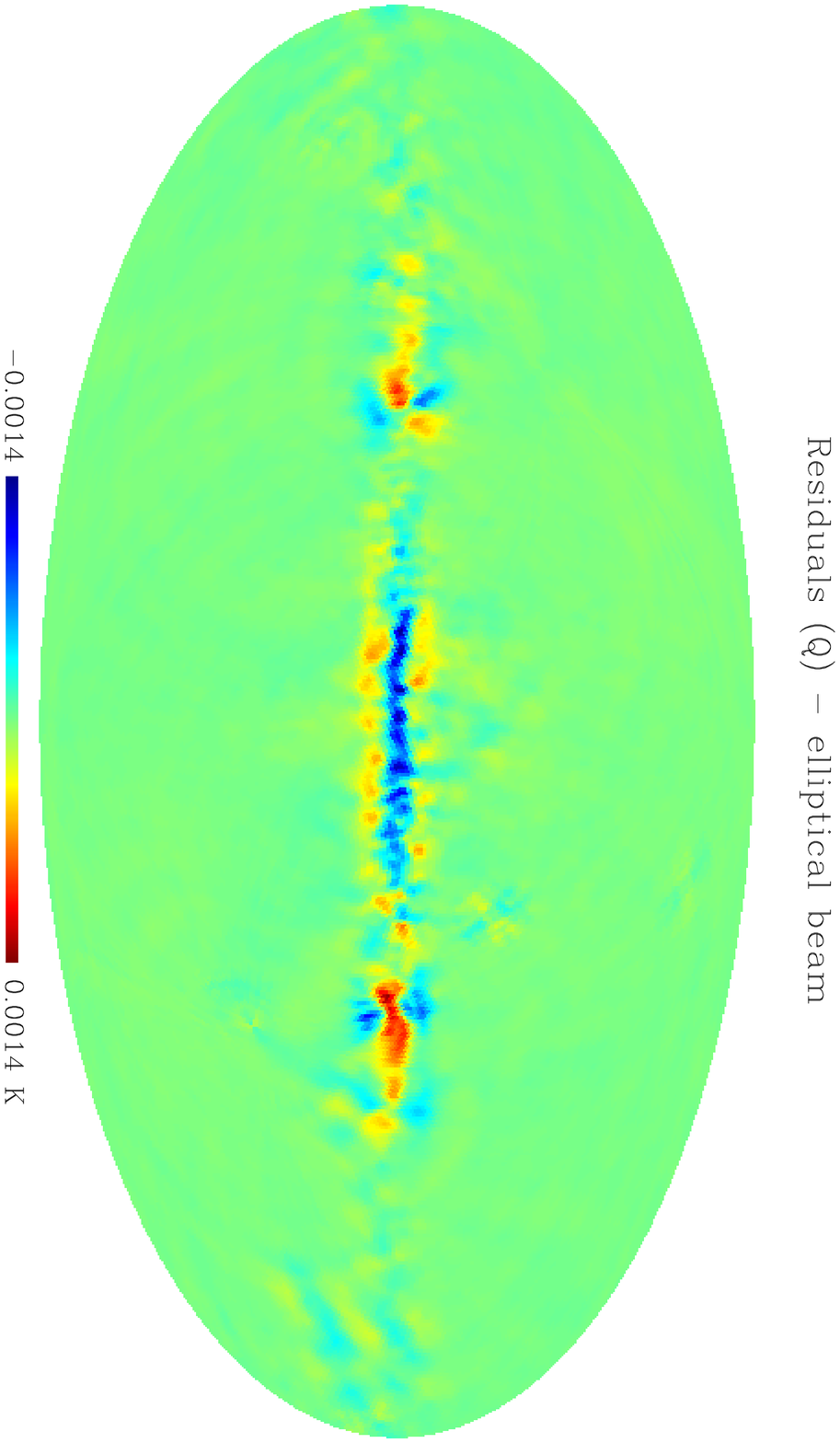}}
%% circular
\put(9,-1.5){\includegraphics{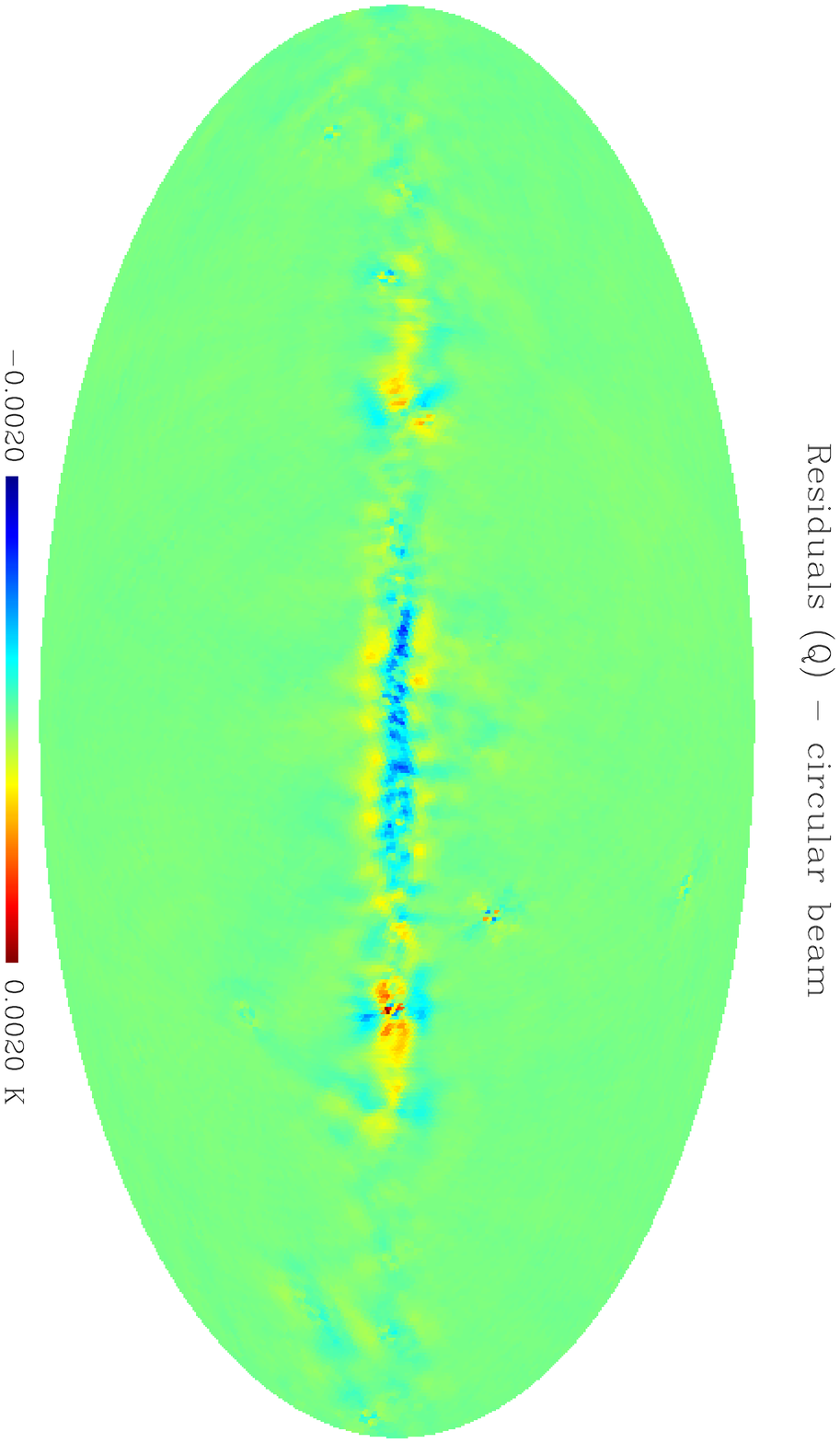}}
\end{picture}
\end{center}
\caption[]{As for Figure~\ref{inhouse_resT_fig} but with the $a_{\ell m}^E$ and
  $a_{\ell m}^B$ multipoles sythesised onto a HEALPix map for the Stokes parameter Q.}
\label{inhouse_resQ_fig}
\end{figure}

\begin{figure}[ht]
\begin{center}
\setlength{\unitlength}{1cm}
\begin{picture}(8,22)(0,0)
%% input
\put(9,15){\includegraphics{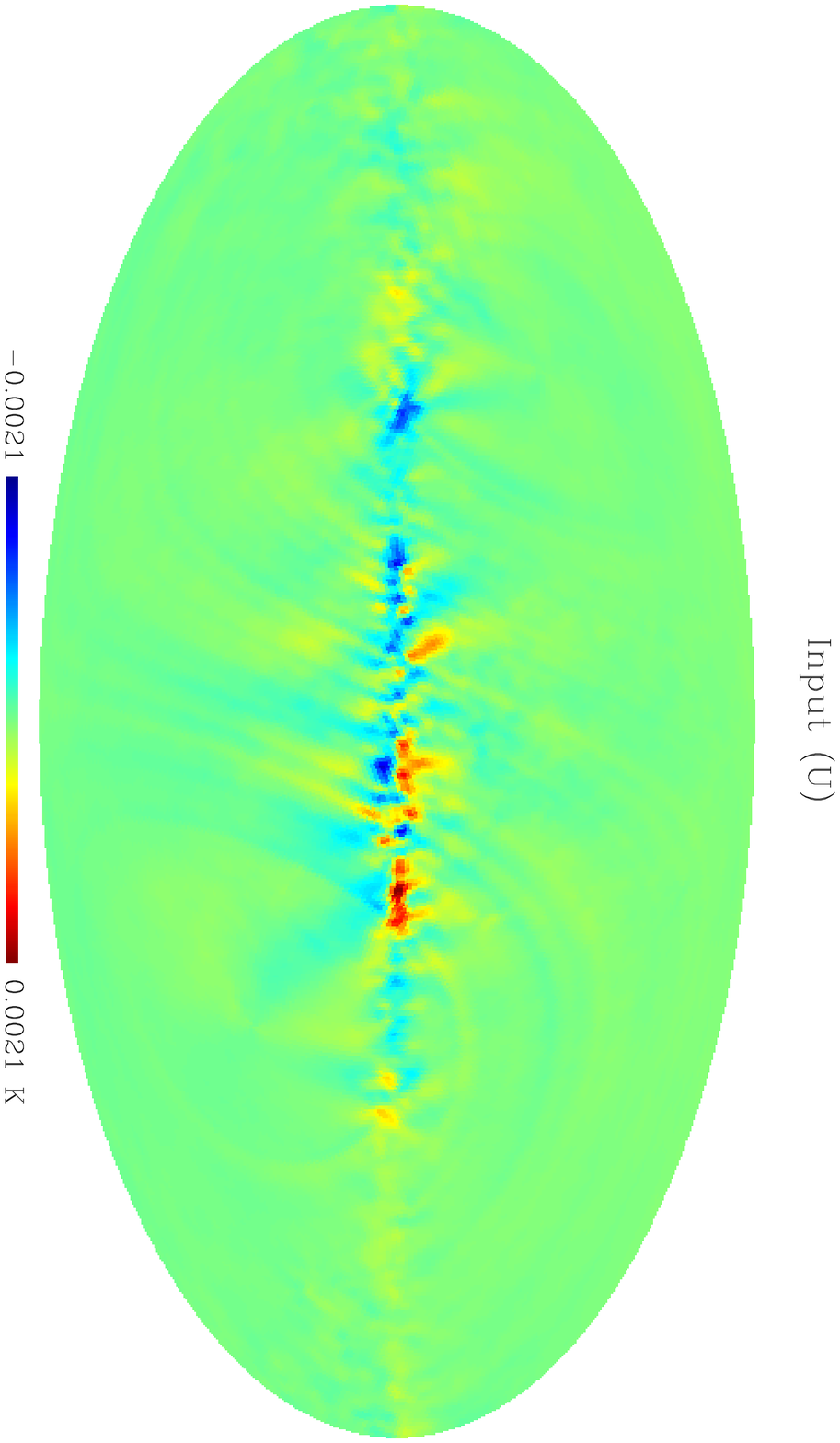}}
%% true
\put(9,9.5){\includegraphics{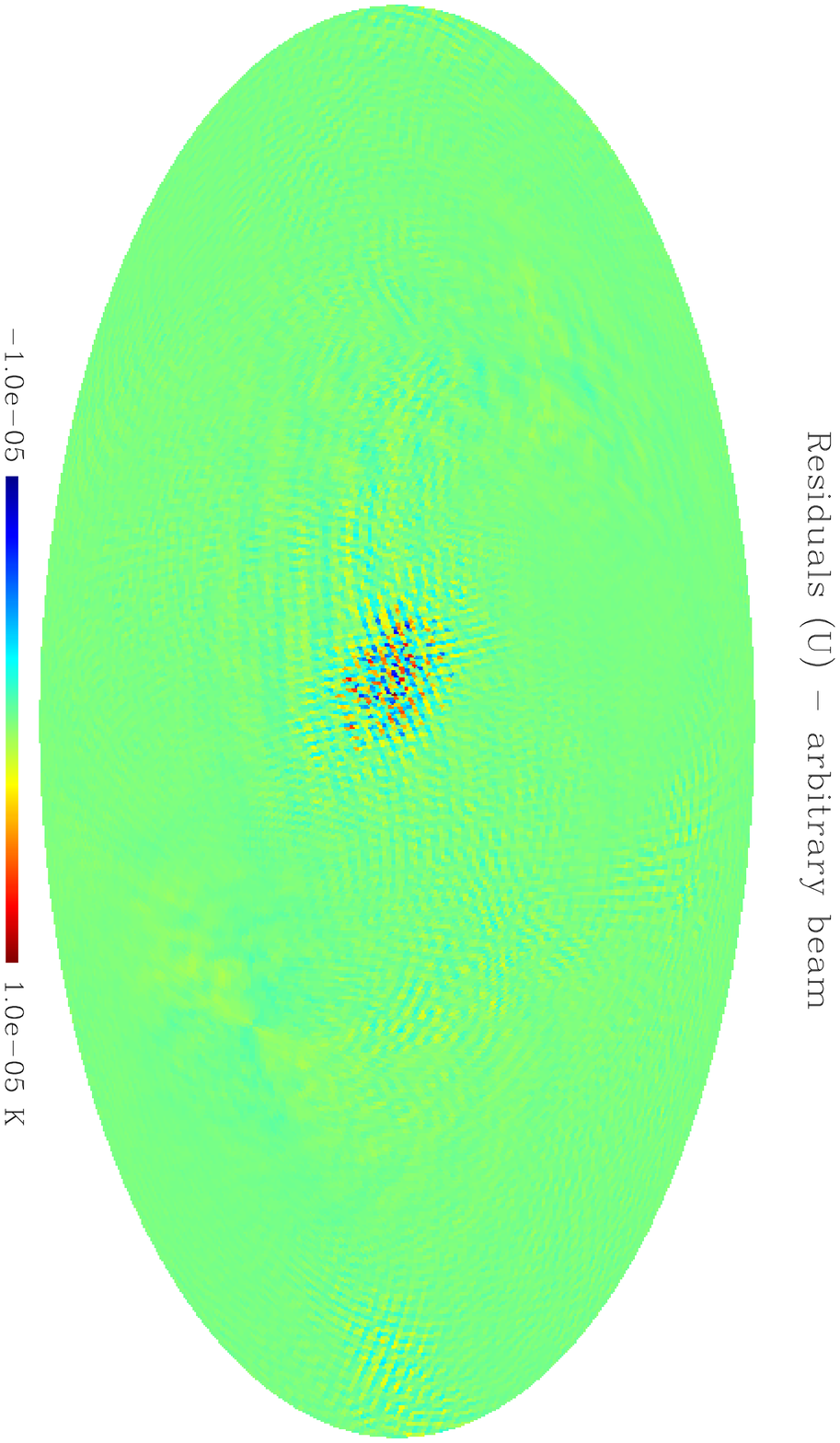}}
%% elliptical
\put(9,4){\includegraphics{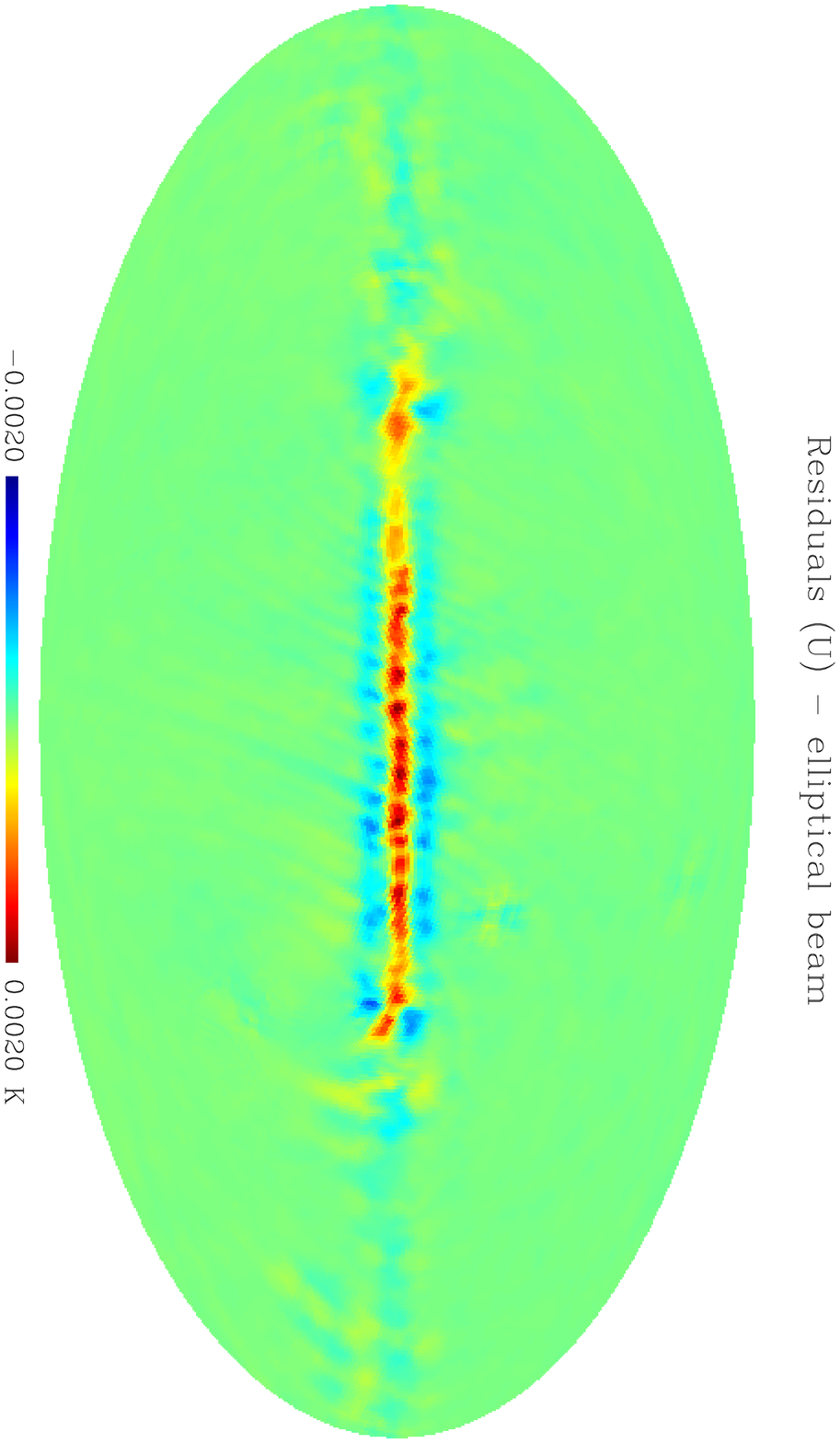}}
%% circular
\put(9,-1.5){\includegraphics{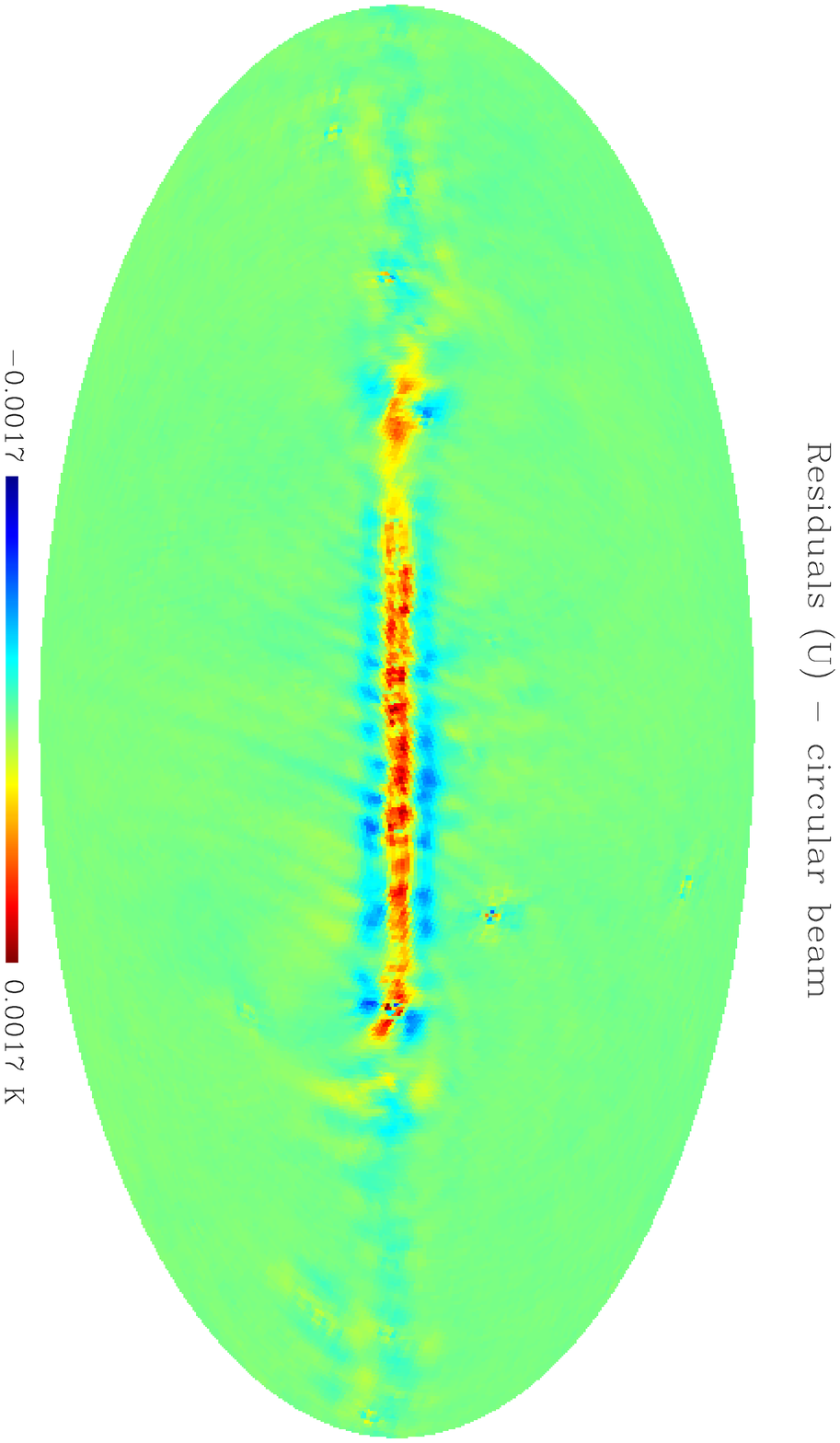}}
\end{picture}
\end{center}
\caption[]{As for Figure~\ref{inhouse_resT_fig} but with the $a_{\ell m}^E$ and
  $a_{\ell m}^B$ multipoles sythesised onto a HEALPix map for the Stokes parameter U.}
\label{inhouse_resU_fig}
\end{figure}

The complex-beam simulation was analysed, using the iterative approach described in Section~\ref{approx_section}, using the three different beams described in Section~\ref{own_sim} and visually represented in Figure~\ref{mybeams_fig}.  The residual multipoles of these analyses have been synthesised onto $T$, $Q$ and $U$ maps shown in Figures~\ref{inhouse_resT_fig},~\ref{inhouse_resQ_fig}~and~\ref{inhouse_resU_fig}, respectively. In each of these figures the input-sky multipoles, which are the same as those used in the {\it Trieste} simulations described in Section~\ref{trieste_sim}, are convolved with a circularly symmetric beam with a FWHM of $1^{\circ}$  and synthesised onto a map for comparison with the structures seen in the residual  maps. 

\begin{figure}
\begin{center}
\setlength{\unitlength}{1cm}
\begin{picture}(8,19)(0,0)
\put(9,11.25){\includegraphics{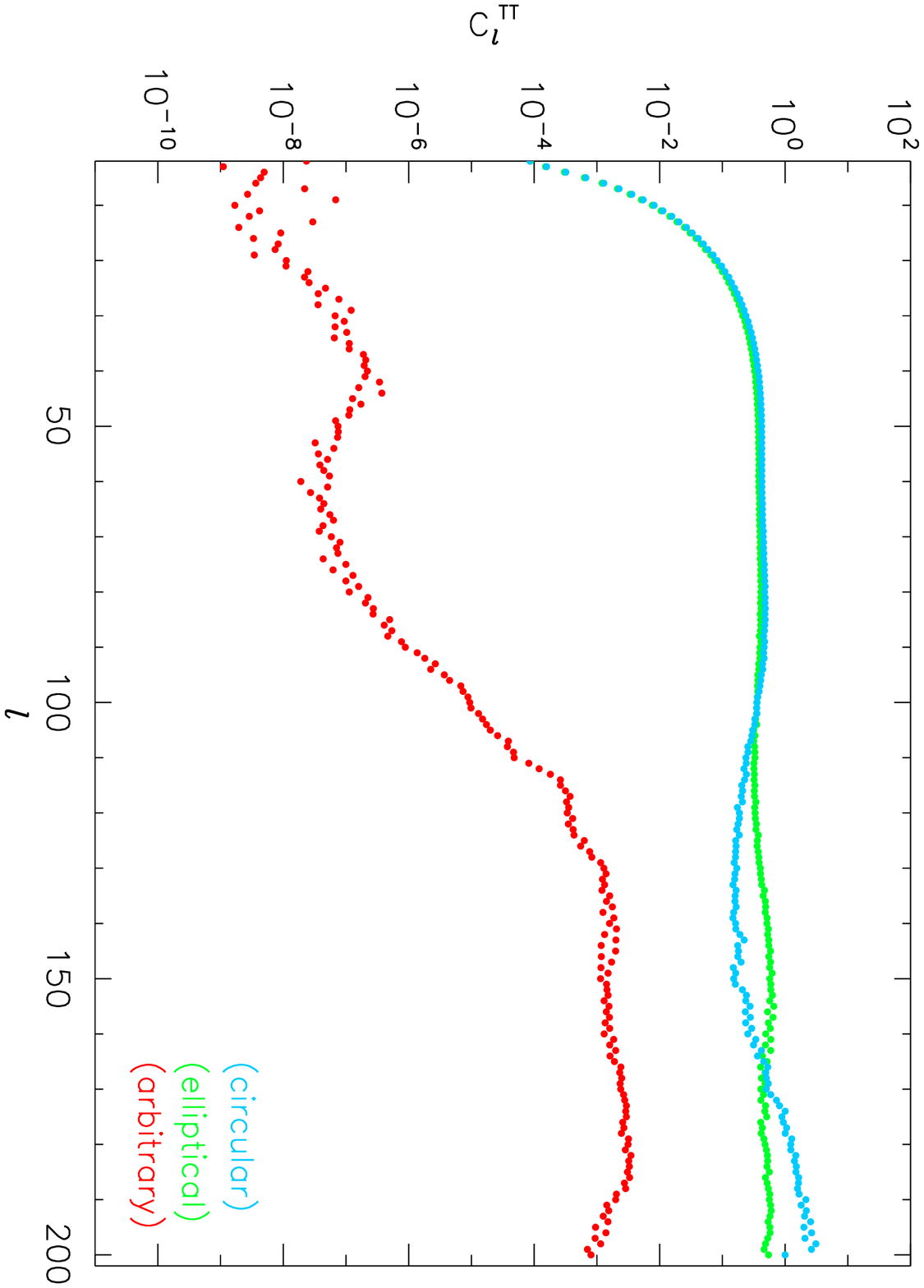}}
\put(9,5){\includegraphics{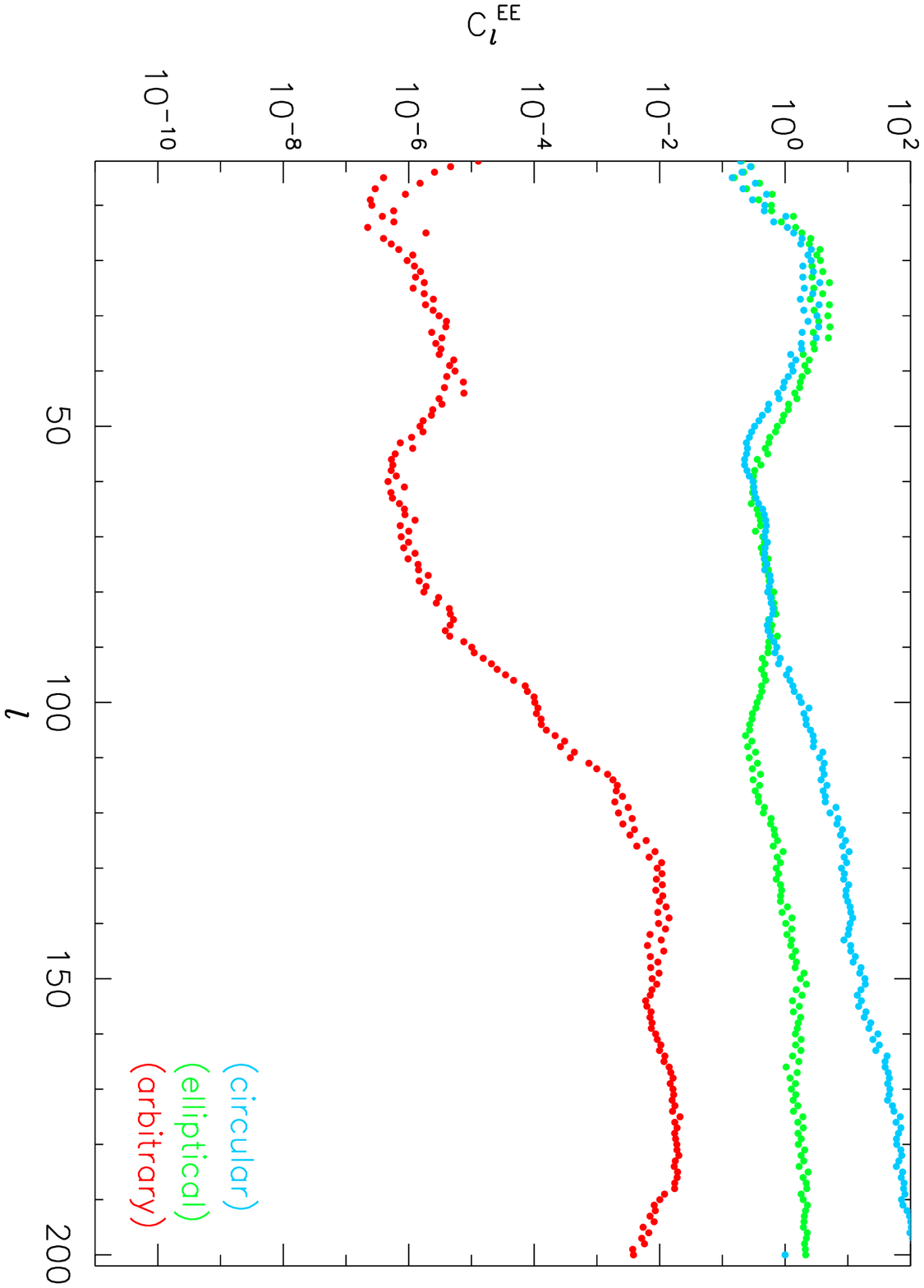}}
\put(9,-1.25){\includegraphics{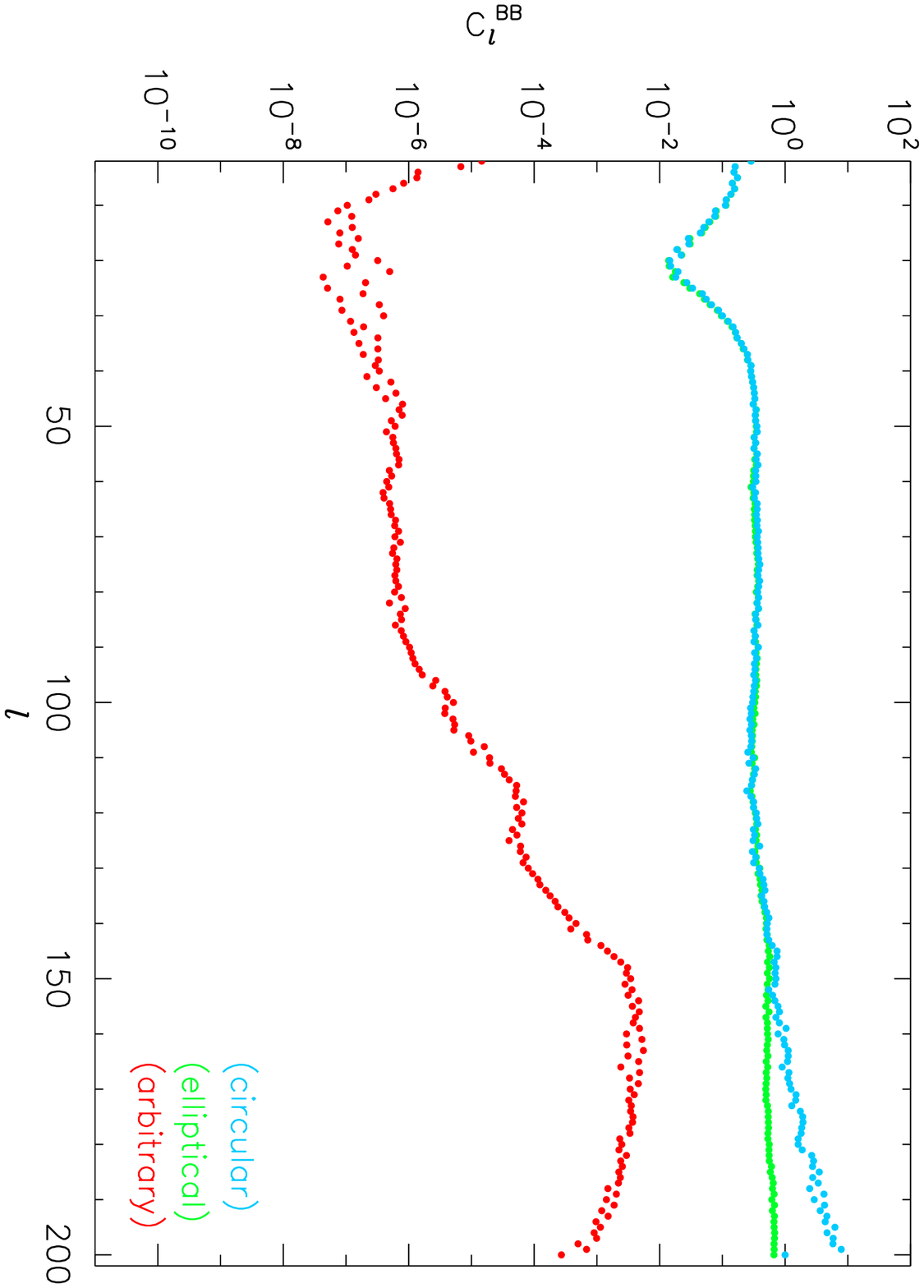}}
\end{picture}
\end{center}
\caption[]{Power spectra of the residuals in $T$ (top), $E$, (middle) and $B$ (bottom) divided by the power spectrum of the input sky. Shown here are the results of the analysis of the complex-beam simulation, by the three different beams. The red, green and blue points show these different analyses using the true, elliptical and circular beams, respectively. As the true beam, including its side-lobes, spans some $15^{\circ}$ all scales are effected by the beam; as seen in the figures in the large difference between the fractional errors of the analysis performed using the true beam and the analyses using the elliptical and circular beams. The fractional errors of the analyses using the elliptical and circular beams, are seen to diverge from the scale at which the difference between these two beams becomes important. }
\label{inhouse_cls_fig}
\end{figure}

An alternate assessment of the performance of the analyses is shown in Figure~\ref{inhouse_cls_fig}. This Figure shows the fractional reconstruction errors in the power spectra, for $T$, $E$, and $B$, in the cases where the true, elliptical or circular beams are used in the deconvolution. It should be noted that, these power spectra do not correspond to the CMB power spectrum, but to a combination of the CMB and all the simulated foregrounds, including point sources.

In Figures~\ref{inhouse_resT_fig}~through~\ref{inhouse_resU_fig}, the maps synthesised from the residual multipoles from the analysis of the complex-beam simulation with the true beams may be seen to contain little power. Additionally, the low level of the fractional errors in the power spectra in Figure~\ref{inhouse_cls_fig}, produced from the analysis using the true beams, also confirms the successful removal of the beam effects from the data. 
 
Figures~\ref{inhouse_resT_fig}~through~\ref{inhouse_resU_fig}, also show how ignoring beam asymmetries may affect the maps. This is seen in the analyses using the elliptical and circular beams, where the residual maps show structure along the galactic plane and regions near compact objects.

%%%%%%%%%%%%%%%%%%%%%%%%%%%%%%%%%%%%%%%%%%%%%%%%%%%%%%%%%%%%%%%%%%%%%%%%%%%
%%Figure for \subsection{Analysis of Complex-Beam Simulations}
%%%%%%%%%%%%%%%%%%%%%%%%%%%%%%%%%%%%%%%%%%%%%%%%%%%%%%%%%%%%%%%%%%%%%%%%%%%

%%%%%%%%%%%%%%%%%%%%%%%%%%%%%%%%%%%%%%%%%%%%%%%%%%%%%%%%%%%%%%%%%%%%%%%%%%%
%% Figure with residual maps for in house simulations -- T
%%%%%%%%%%%%%%%%%%%%%%%%%%%%%%%%%%%%%%%%%%%%%%%%%%%%%%%%%%%%%%%%%%%%%%%%%%%

%%%%%%%%%%%%%%%%%%%%%%%%%%%%%%%%%%%%%%%%%%%%%%%%%%%%%%%%%%%%%%%%%%%%%%%%%%%
%% Figure with residual maps for in house simulations -- Q
%%%%%%%%%%%%%%%%%%%%%%%%%%%%%%%%%%%%%%%%%%%%%%%%%%%%%%%%%%%%%%%%%%%%%%%%%%%

%%%%%%%%%%%%%%%%%%%%%%%%%%%%%%%%%%%%%%%%%%%%%%%%%%%%%%%%%%%%%%%%%%%%%%%%%%%
%% Figure with residual maps for in house simulations -- U
%%%%%%%%%%%%%%%%%%%%%%%%%%%%%%%%%%%%%%%%%%%%%%%%%%%%%%%%%%%%%%%%%%%%%%%%%%%

%%%%%%%%%%%%%%%%%%%%%%%%%%%%%%%%%%%%%%%%%%%%%%%%%%%%%%%%%%%%%%%%%%%%%%%%%%%
%% cl figure -- in house simulation results
%%%%%%%%%%%%%%%%%%%%%%%%%%%%%%%%%%%%%%%%%%%%%%%%%%%%%%%%%%%%%%%%%%%%%%%%%%%
%%%%%%%%%%%%%%%%%%%%%%%%%%%%%%%%%%%%%%%%%%%%%%%%%%%%%%%%%%%%%%%%%%%%%%%%%%%

%%%%%%%%%%%%%%%%%%%%%%%%%%%%%%%%%%%%%%%%%%%%%%%%%%%%%%%%%%%%%%%%%%%%%%%%%%%
%% Figures for \subsection{Analysis of {\it Trieste} data}
%%%%%%%%%%%%%%%%%%%%%%%%%%%%%%%%%%%%%%%%%%%%%%%%%%%%%%%%%%%%%%%%%%%%%%%%%%%

%%%%%%%%%%%%%%%%%%%%%%%%%%%%%%%%%%%%%%%%%%%%%%%%%%%%%%%%%%%%%%%%%%%%%%%%%%%
%% Figure with maps for the case of no Pt & SZ sources -- T
%%%%%%%%%%%%%%%%%%%%%%%%%%%%%%%%%%%%%%%%%%%%%%%%%%%%%%%%%%%%%%%%%%%%%%%%%%%
\begin{figure}[ht]
\begin{center}
\setlength{\unitlength}{1cm}
\begin{picture}(8,16.5)(0,0)
%% true sky row
\put(9,9){\includegraphics{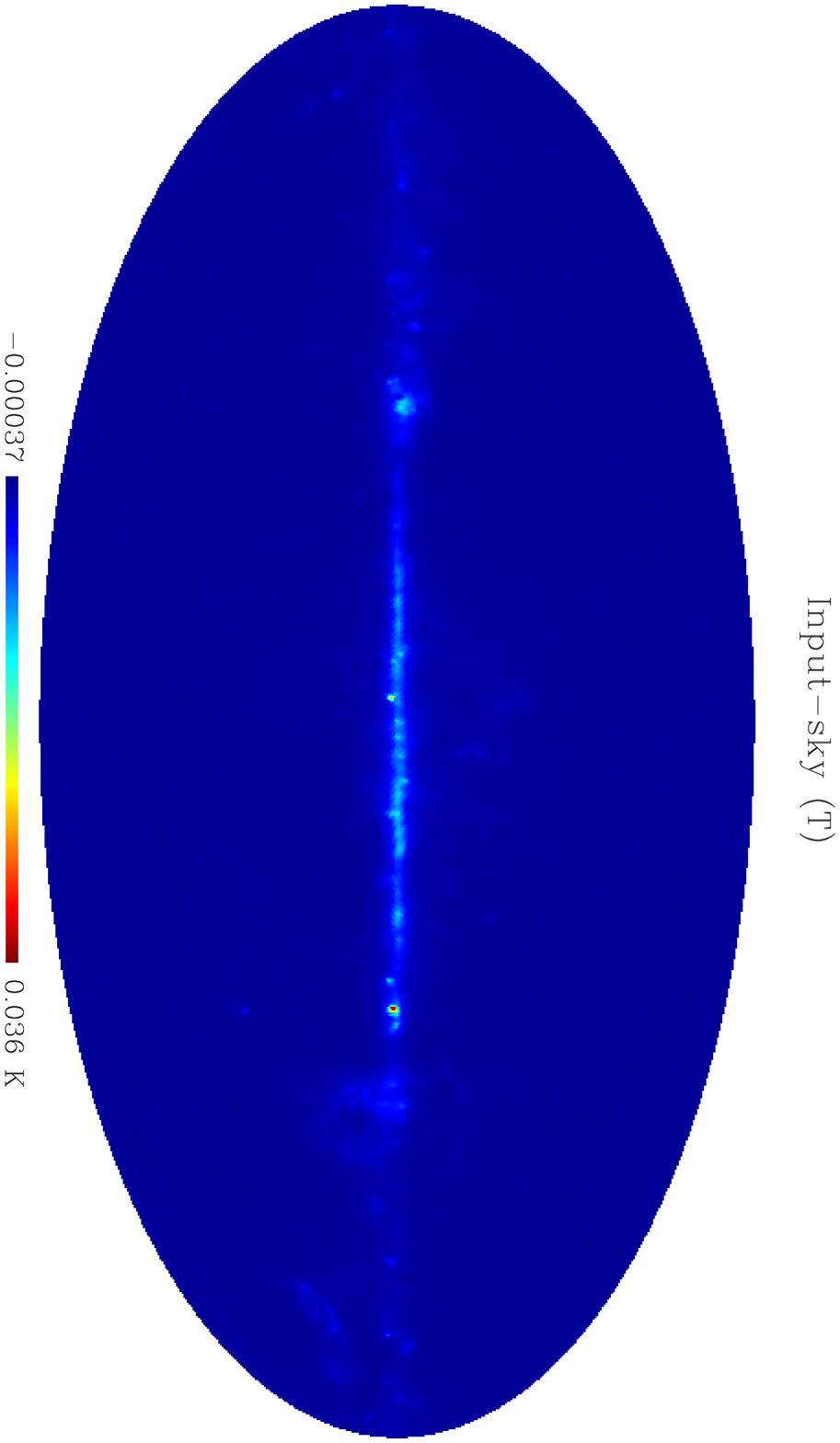}}
%% map row
\put(9,3.75){\includegraphics{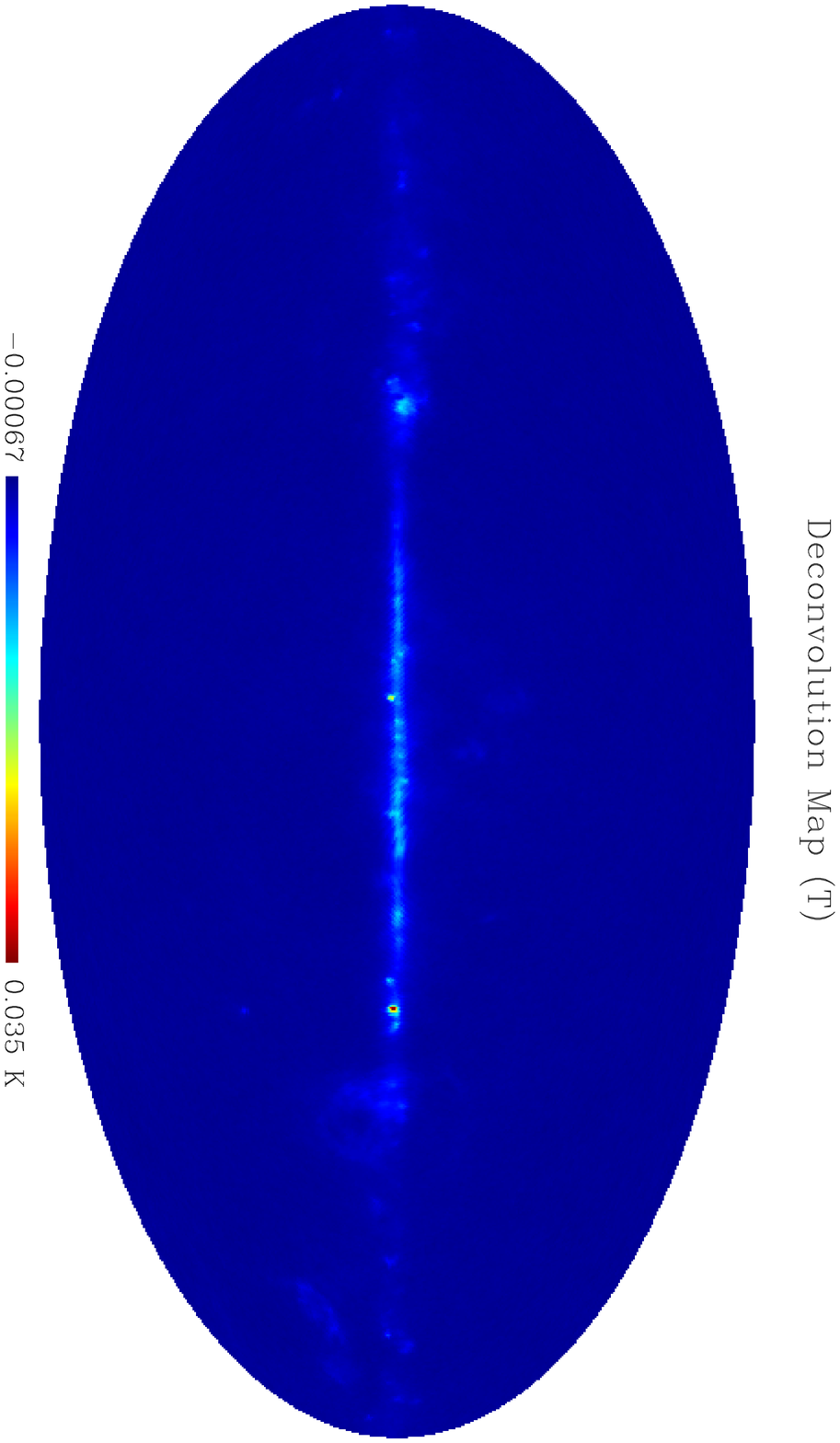}}
%% residuals row
\put(9,-1.5){\includegraphics{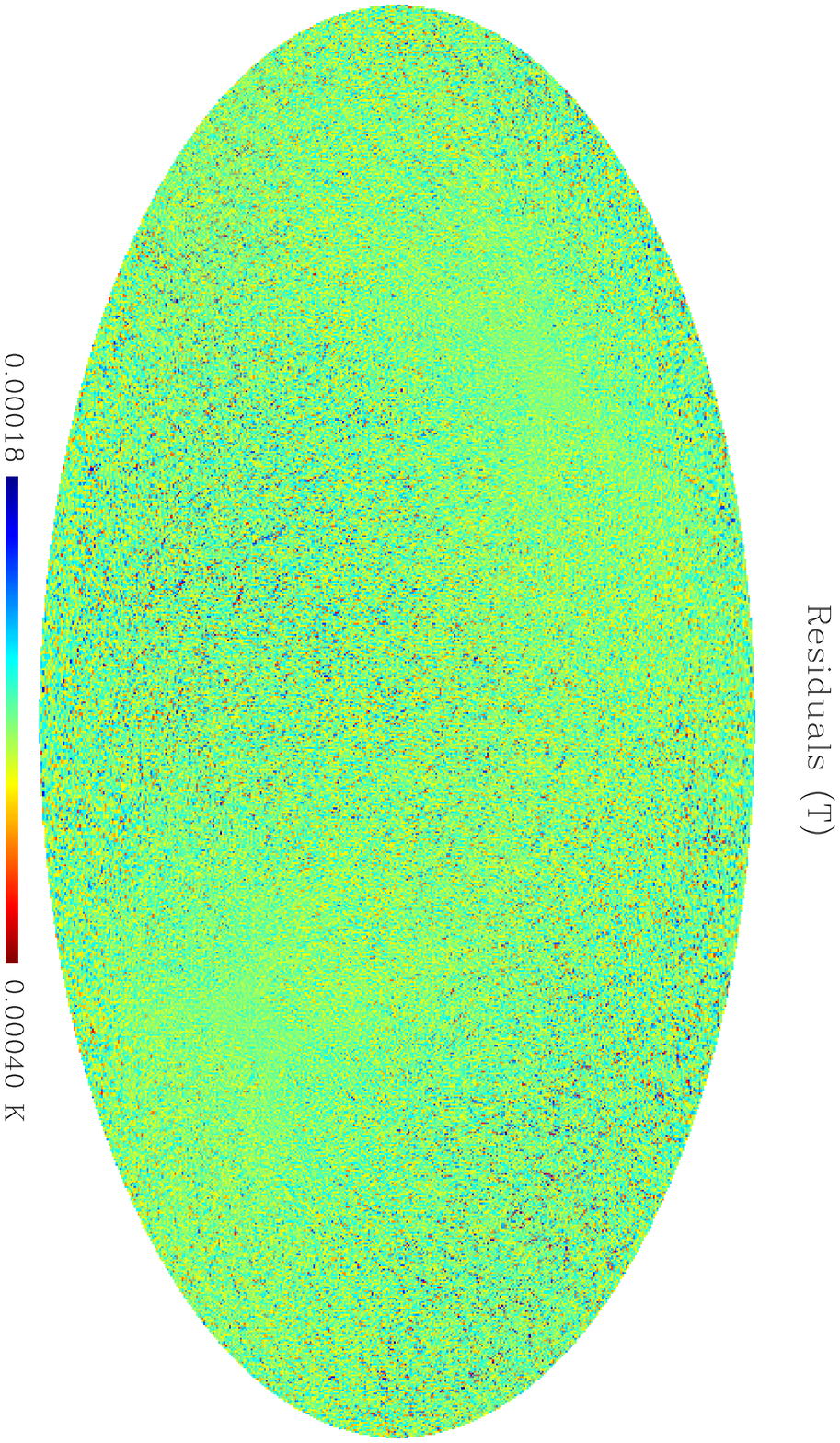}}
\end{picture}
\end{center}
\caption[]{Top: Temperature multipoles, $a^T_{\ell \it m}$, as input to the {\it Trieste} simulations, curtailed at $\ell=400$, and synthesised onto a HEALPix map of $\rm{ N_{side}}$=128. These $a_{\ell \it m}$ include contributions from the CMB, synchrotron, free-free, and dust. Middle:  $a^T_{\ell \it m}$ recovered from the analysis, up to $\ell_{max}=400$, of the corresponding {\it Trieste} simulations which included the effects of sampling, elliptical beams and $1/f$ noise. These  $a^T_{\ell \it m}$ are again synthesised onto a HEALPix map of $\rm{ N_{side}}$=128. It should be noted that the simulations analysed here contain power above $\ell=400$ and it is only for comparison purposes that we curtail the input temperature multipoles to the value of $\ell_{max}$ to which the recovery proceeded. Bottom: Residuals between the input-sky and the recovered  $a^T_{\ell \it m}$.} These residuals are consistent with noise, the visible pattern is due to the scanning strategy.
\label{noptsT_fig}
\end{figure}

%%%%%%%%%%%%%%%%%%%%%%%%%%%%%%%%%%%%%%%%%%%%%%%%%%%%%%%%%%%%%%%%%%%%%%%%%%%
%% Figure with maps for the case of no Pt & SZ sources -- Q
%%%%%%%%%%%%%%%%%%%%%%%%%%%%%%%%%%%%%%%%%%%%%%%%%%%%%%%%%%%%%%%%%%%%%%%%%%%
\begin{figure}[ht]
\begin{center}
\setlength{\unitlength}{1cm}
\begin{picture}(8,16.5)(0,0)
%% true sky row
\put(9,9){\includegraphics{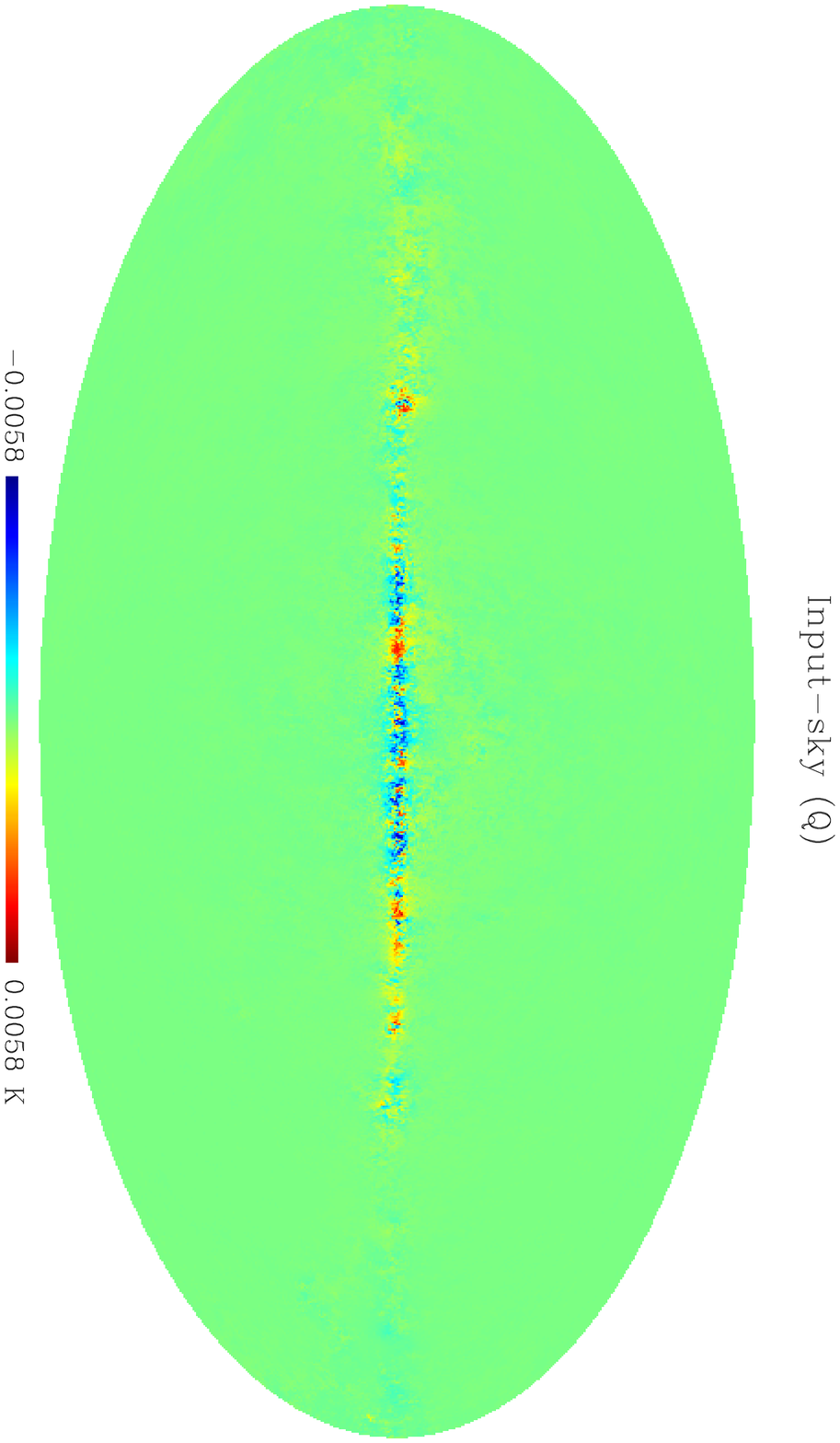}}
%% map row
\put(9,3.75){\includegraphics{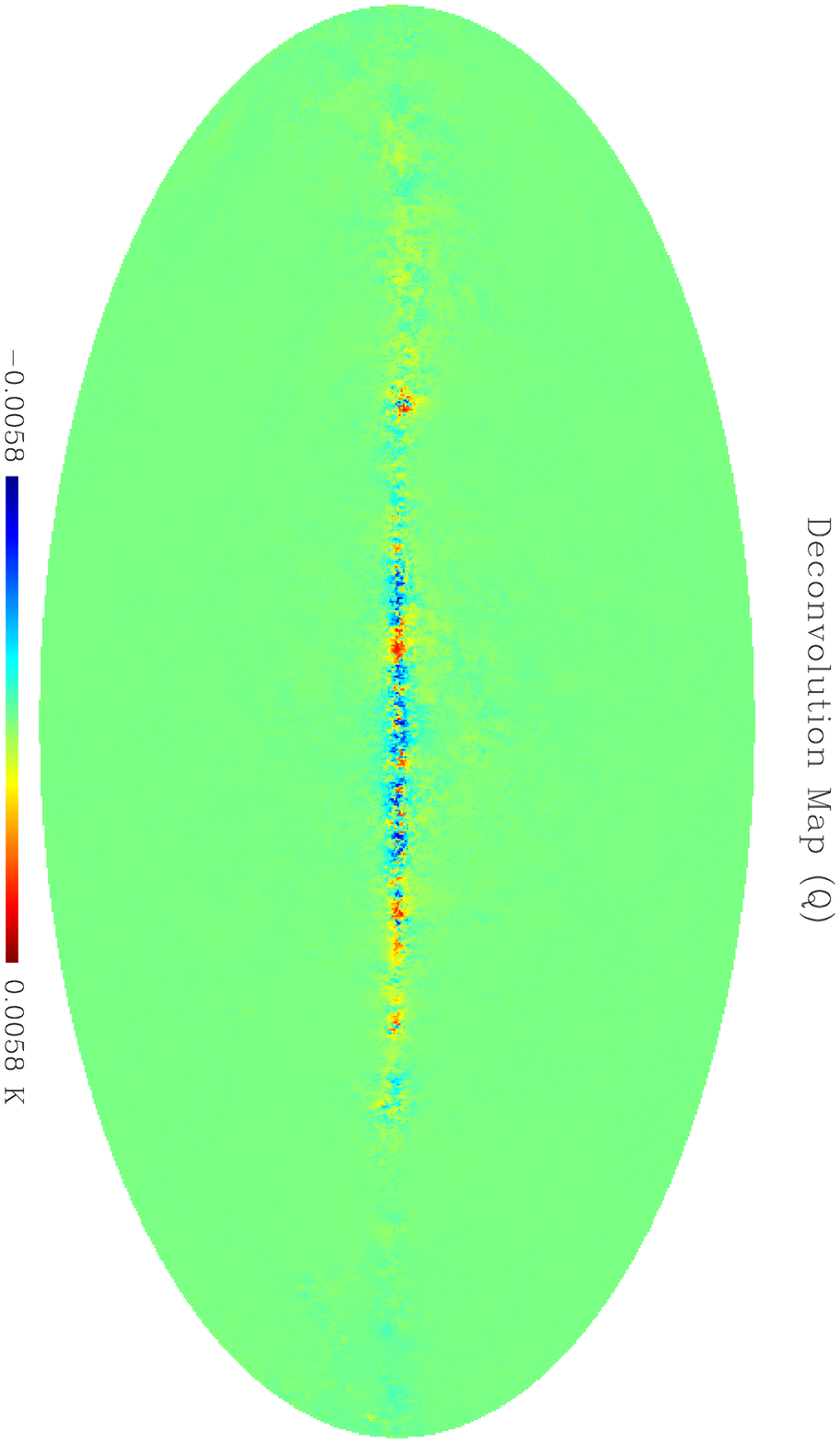}}
%% residuals row
\put(9,-1.5){\includegraphics{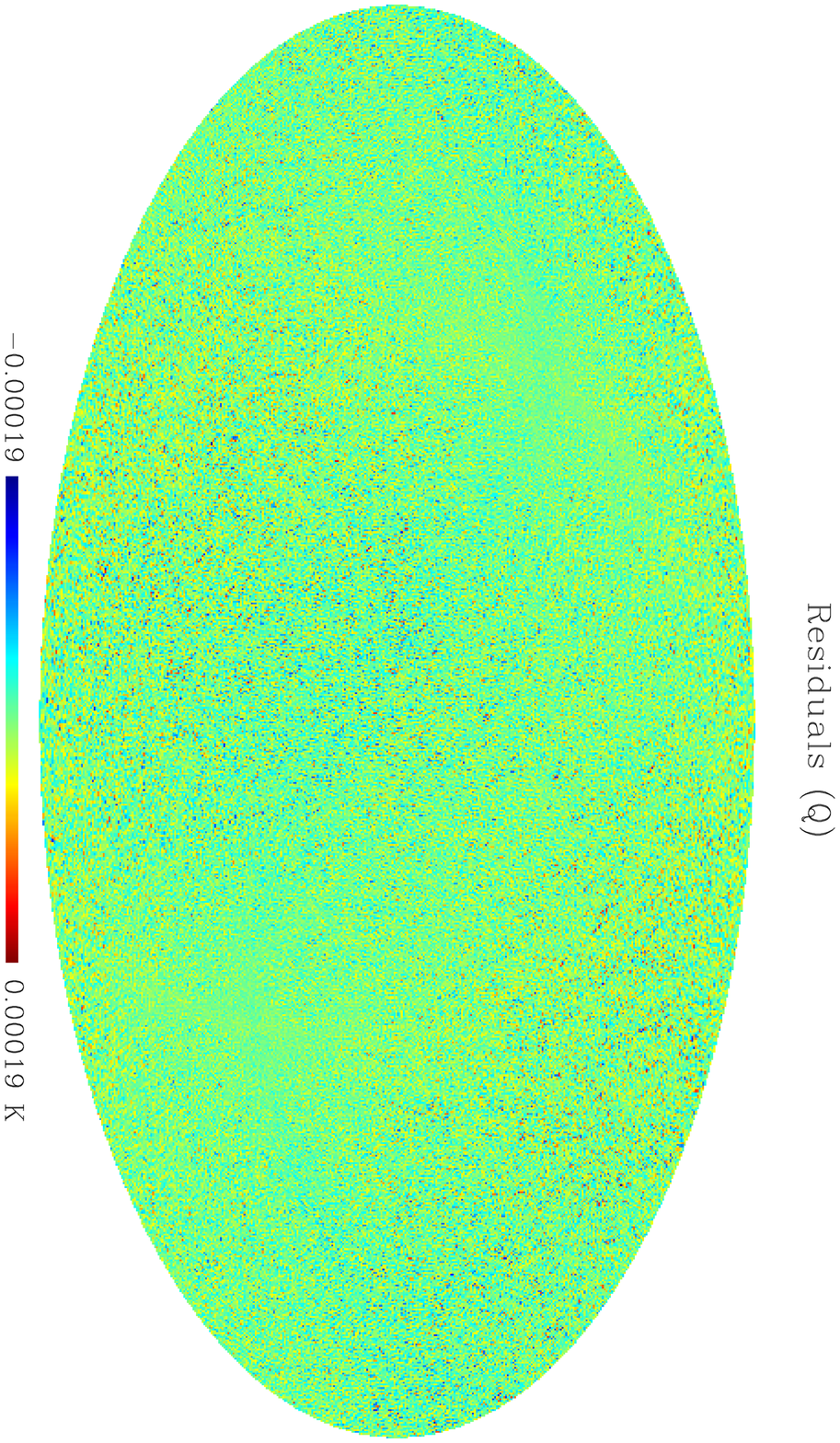}}
\end{picture}
\end{center}
\caption[]{As for Figure~\ref{noptsT_fig} but with the $a_{\ell m}^E$ and
  $a_{\ell m}^B$ multipole sythesised onto a HEALPix map for the Stokes parameter Q.}
\label{noptsQ_fig}
\end{figure}
%%%%%%%%%%%%%%%%%%%%%%%%%%%%%%%%%%%%%%%%%%%%%%%%%%%%%%%%%%%%%%%%%%%%%%%%%%%

%%%%%%%%%%%%%%%%%%%%%%%%%%%%%%%%%%%%%%%%%%%%%%%%%%%%%%%%%%%%%%%%%%%%%%%%%%%
%% Figure with maps for the case of no Pt & SZ sources -- U
%%%%%%%%%%%%%%%%%%%%%%%%%%%%%%%%%%%%%%%%%%%%%%%%%%%%%%%%%%%%%%%%%%%%%%%%%%%
\begin{figure}[ht]
\begin{center}
\setlength{\unitlength}{1cm}
\begin{picture}(8,16.5)(0,0)
%% true sky row
\put(9,9){\includegraphics{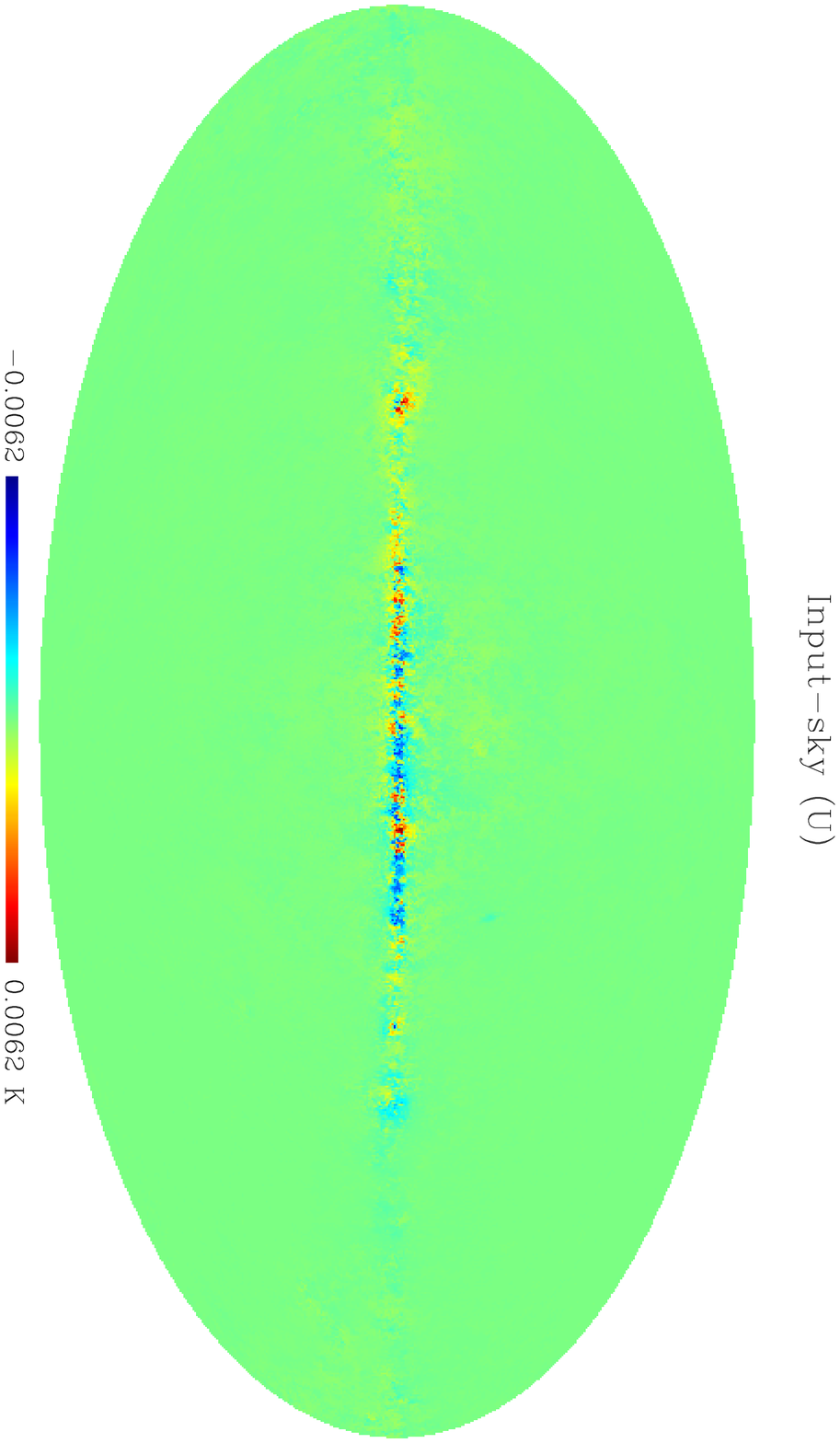}}
%% map row
\put(9,3.75){\includegraphics{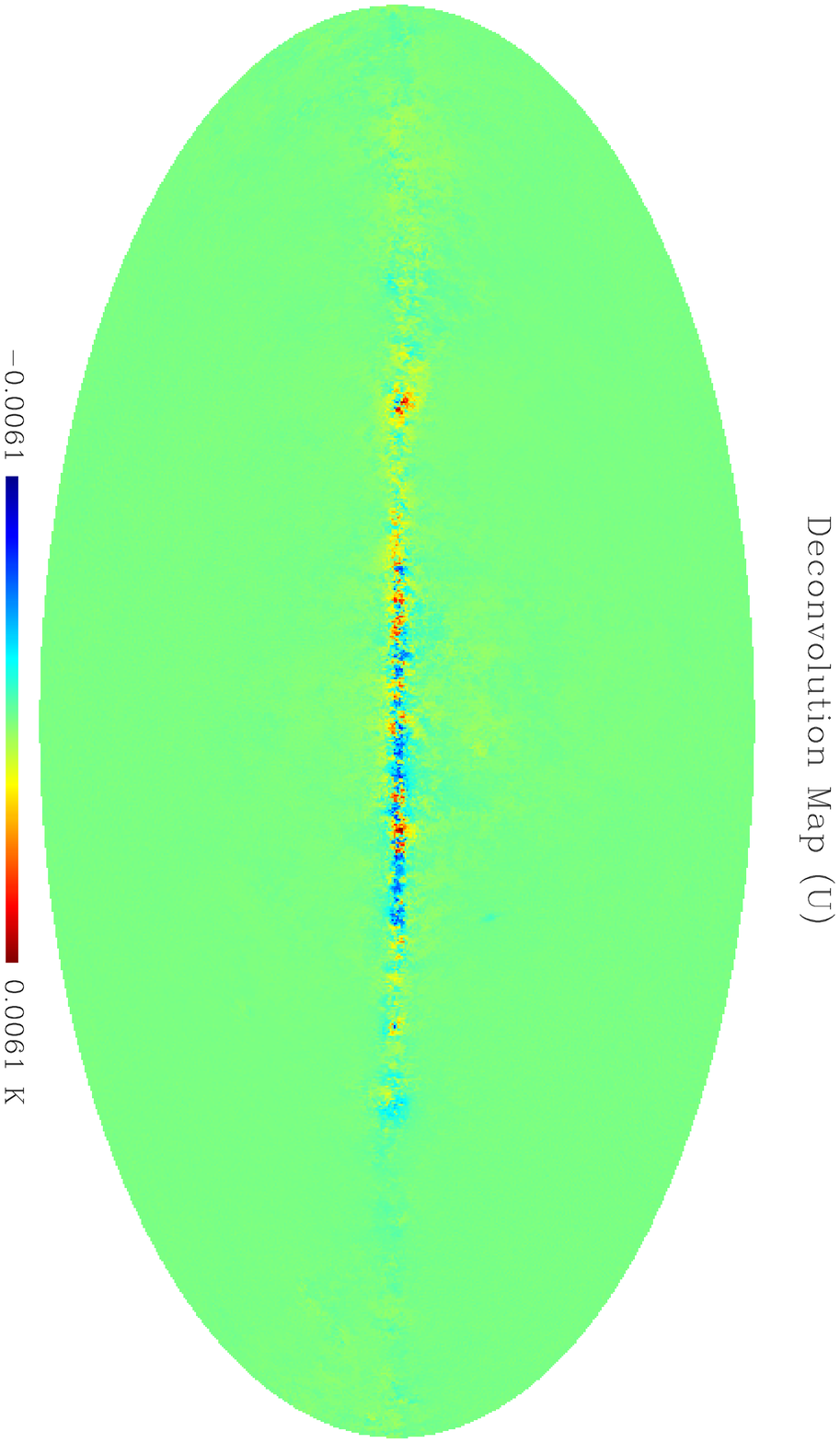}}
%% residuals row
\put(9,-1.5){\includegraphics{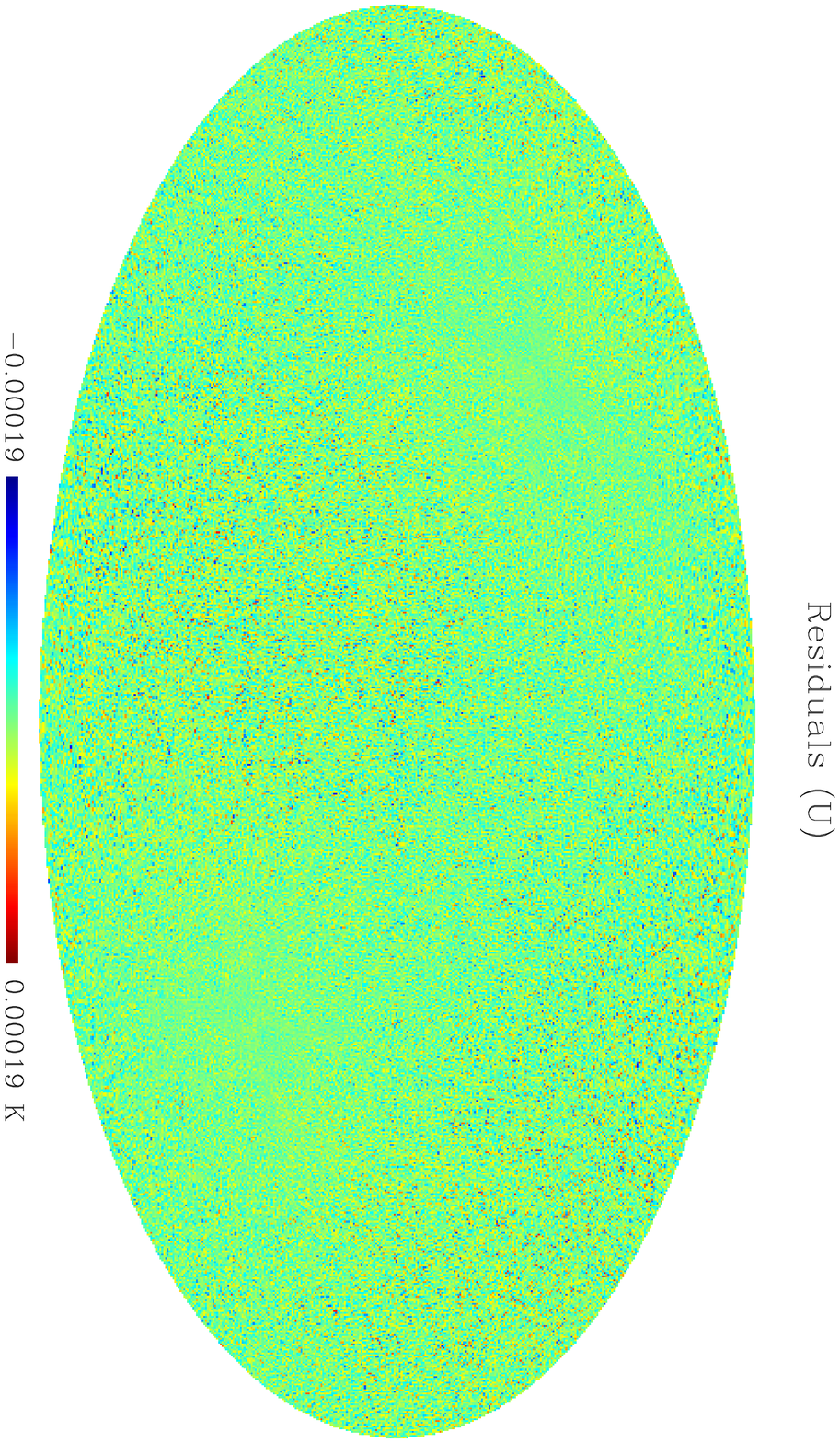}}
\end{picture}
\end{center}
\caption[]{As for Figure~\ref{noptsQ_fig} but for the Stokes parameter U.}
\label{noptsU_fig}
\end{figure}
%%%%%%%%%%%%%%%%%%%%%%%%%%%%%%%%%%%%%%%%%%%%%%%%%%%%%%%%%%%%%%%%%%%%%%%%%%%
%%%%%%%%%%%%%%%%%%%%%%%%%%%%%%%%%%%%%%%%%%%%%%%%%%%%%%%%%%%%%%%%%%%%%%%%%%%

%%%%%%%%%%%%%%%%%%%%%%%%%%%%%%%%%%%%%%%%%%%%%%%%%%%%%%%%%%%%%%%%%%%%%%%%%%%
%% comparison of residuals
%%  -- T
%%%%%%%%%%%%%%%%%%%%%%%%%%%%%%%%%%%%%%%%%%%%%%%%%%%%%%%%%%%%%%%%%%%%%%%%%%%
\begin{figure}
\begin{center}
\setlength{\unitlength}{1cm}
\begin{picture}(8,16.5)(0,0)
%% map row
\put(9,9){\includegraphics{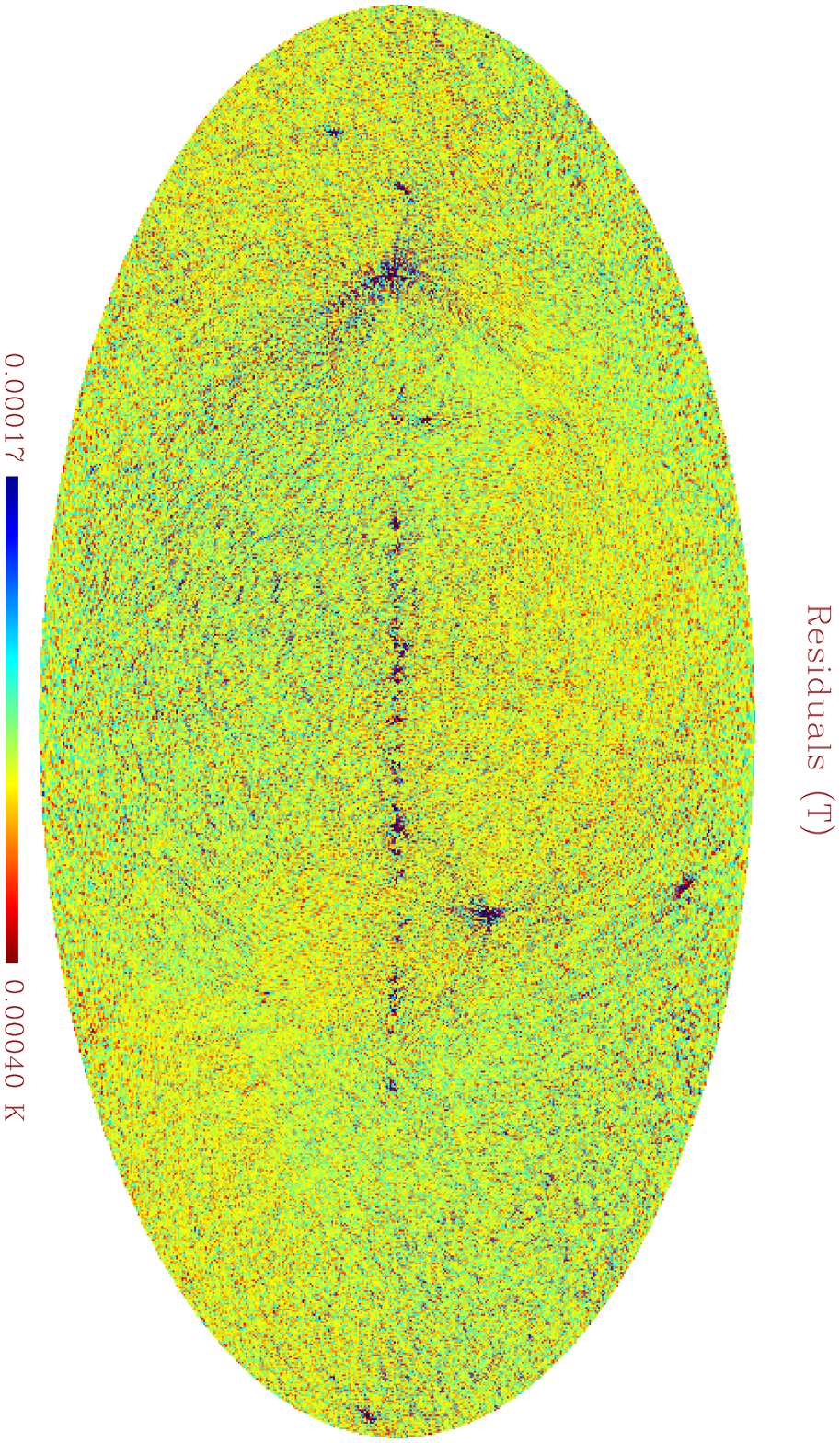}}
%% residuals row
\put(9,3.75){\includegraphics{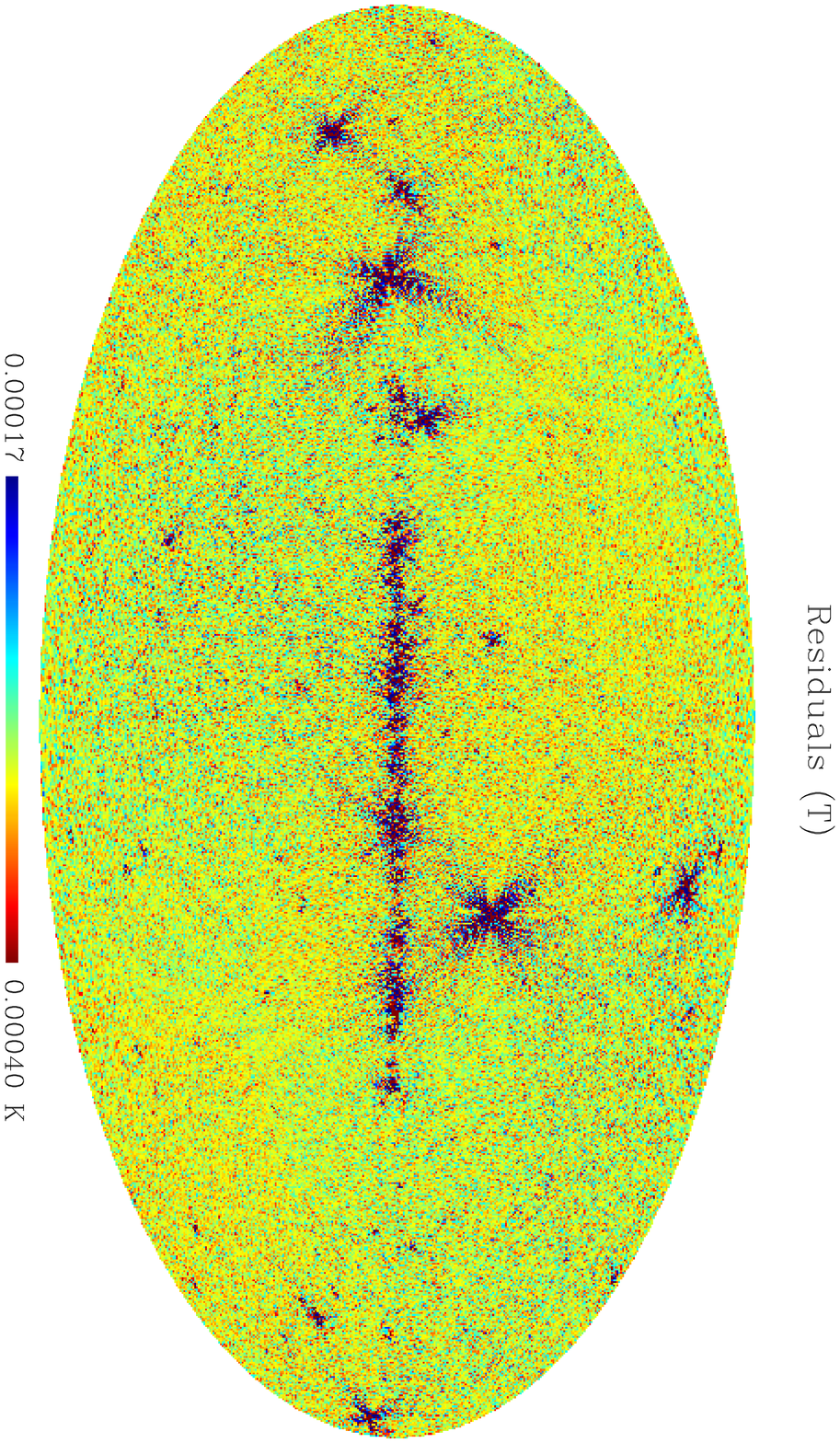}}
%% asym-sym row
\put(9,-1.5){\includegraphics{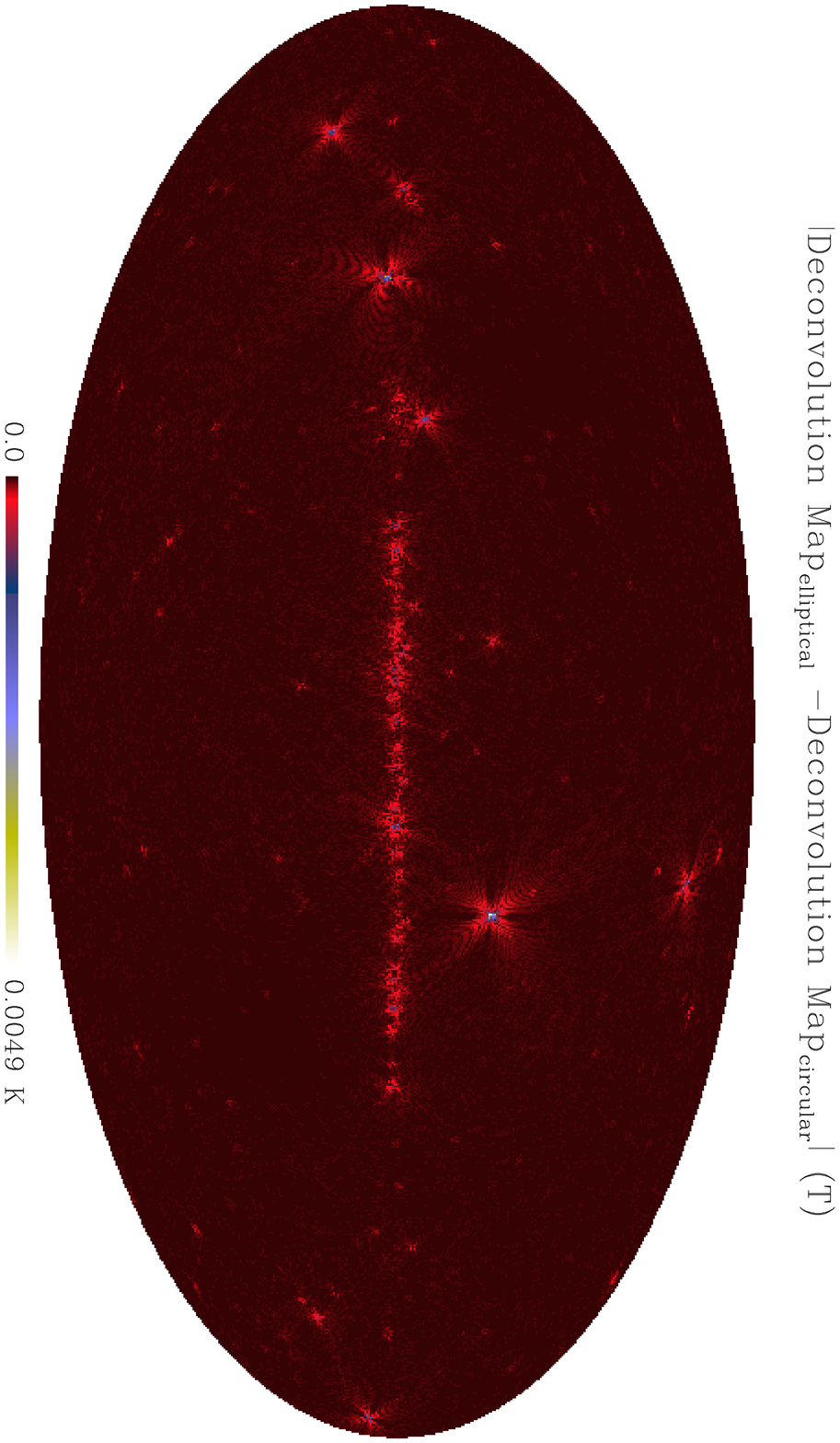}}
\end{picture}
\end{center}
\caption[]{Top: Residuals between the input-sky, including SZ and point
  sources, and the recovered map, plotted over the same range as the residuals
  found when SZ and point sources were not included in the input-sky. Middle: Residuals between the input-sky, including SZ and point sources, and the map recovered using the circular beams, plotted over the same range as the top panel. Bottom: Absolute differences between the maps recovered using the elliptical and circular beams, in the case when SZ and point sources are included. }
\label{comp_Tresiduals_fig}
\end{figure}
%%%%%%%%%%%%%%%%%%%%%%%%%%%%%%%%%%%%%%%%%%%%%%%%%%%%%%%%%%%%%%%%%%%%%%%%%%%

%%%%%%%%%%%%%%%%%%%%%%%%%%%%%%%%%%%%%%%%%%%%%%%%%%%%%%%%%%%%%%%%%%%%%%%%%%%
%%  comparison of residuals
%%  -- Q
%%%%%%%%%%%%%%%%%%%%%%%%%%%%%%%%%%%%%%%%%%%%%%%%%%%%%%%%%%%%%%%%%%%%%%%%%%%
\begin{figure}
\begin{center}
\setlength{\unitlength}{1cm}
\begin{picture}(8,11)(0,0)
%% map row
\put(-0.5,4.75){\includegraphics{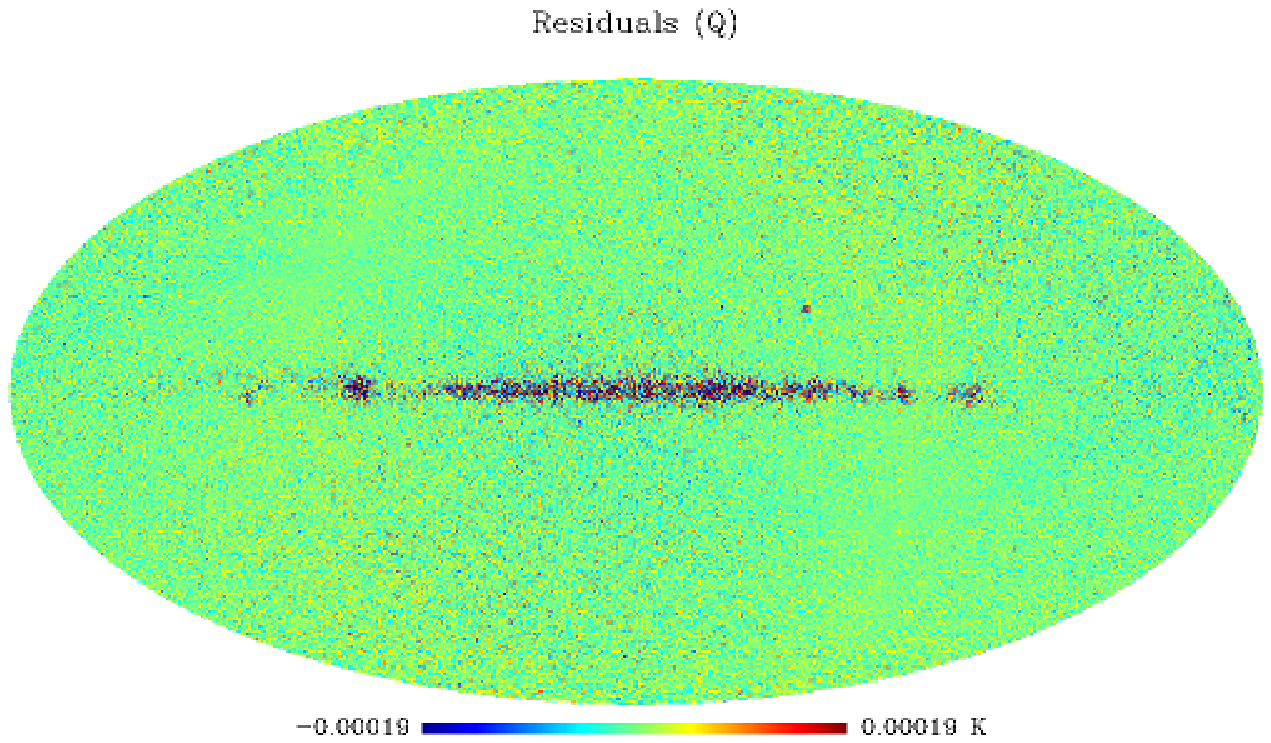}}
%% residuals row
\put(-0.5,-0.5){\includegraphics{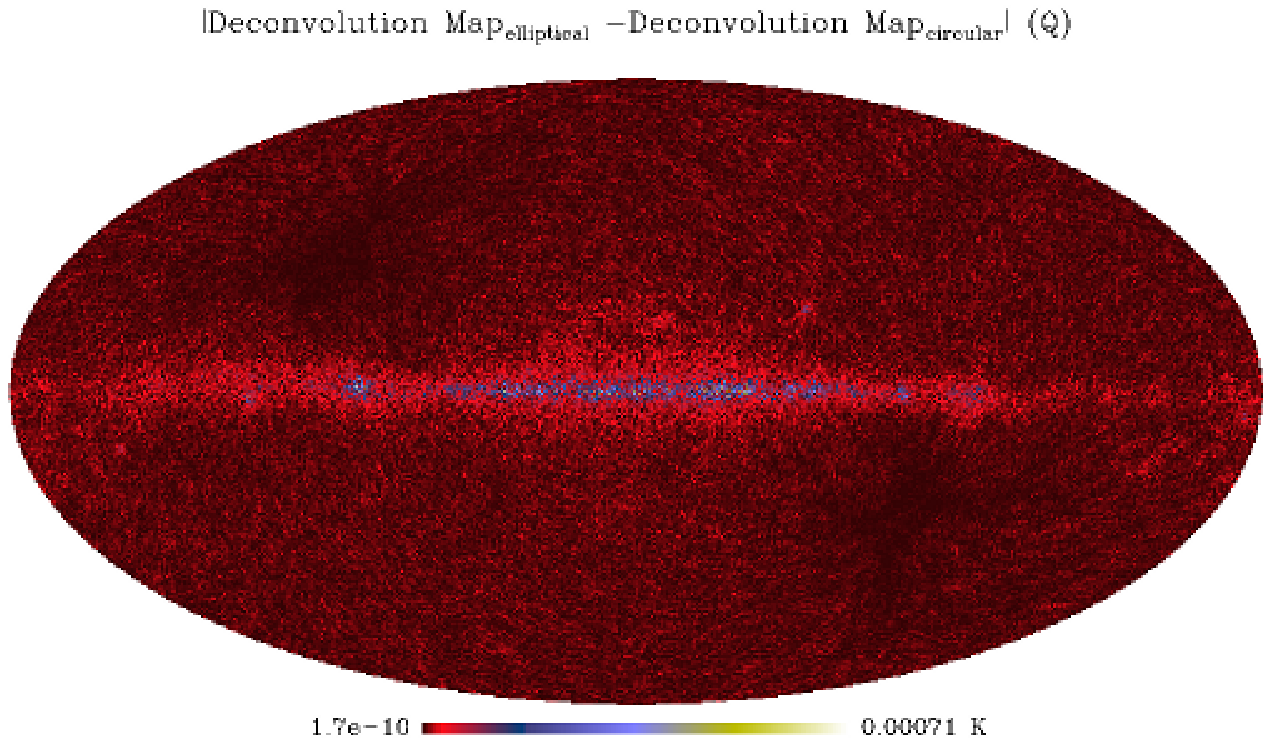}}
\end{picture}
\end{center}
\caption[]{Top: Residuals, for the Stokes Q parameter, between the input-sky including SZ and point sources and the map recovered using the circular beams, plotted over the same range as the residuals found for this parameter when the input-sky does not include SZ and point sources. Bottom: Absolute differences between the maps recovered using elliptical and circular beams.   }
\label{compQresiduals_fig}
\end{figure}
%%%%%%%%%%%%%%%%%%%%%%%%%%%%%%%%%%%%%%%%%%%%%%%%%%%%%%%%%%%%%%%%%%%%%%%%%%%

%%%%%%%%%%%%%%%%%%%%%%%%%%%%%%%%%%%%%%%%%%%%%%%%%%%%%%%%%%%%%%%%%%%%%%%%%%%
%%  comparison of residuals
%%  -- U
%%%%%%%%%%%%%%%%%%%%%%%%%%%%%%%%%%%%%%%%%%%%%%%%%%%%%%%%%%%%%%%%%%%%%%%%%%%
\begin{figure}
\begin{center}
\setlength{\unitlength}{1cm}
\begin{picture}(8,11)(0,0)
%% map row
\put(-0.5,4.75){\includegraphics{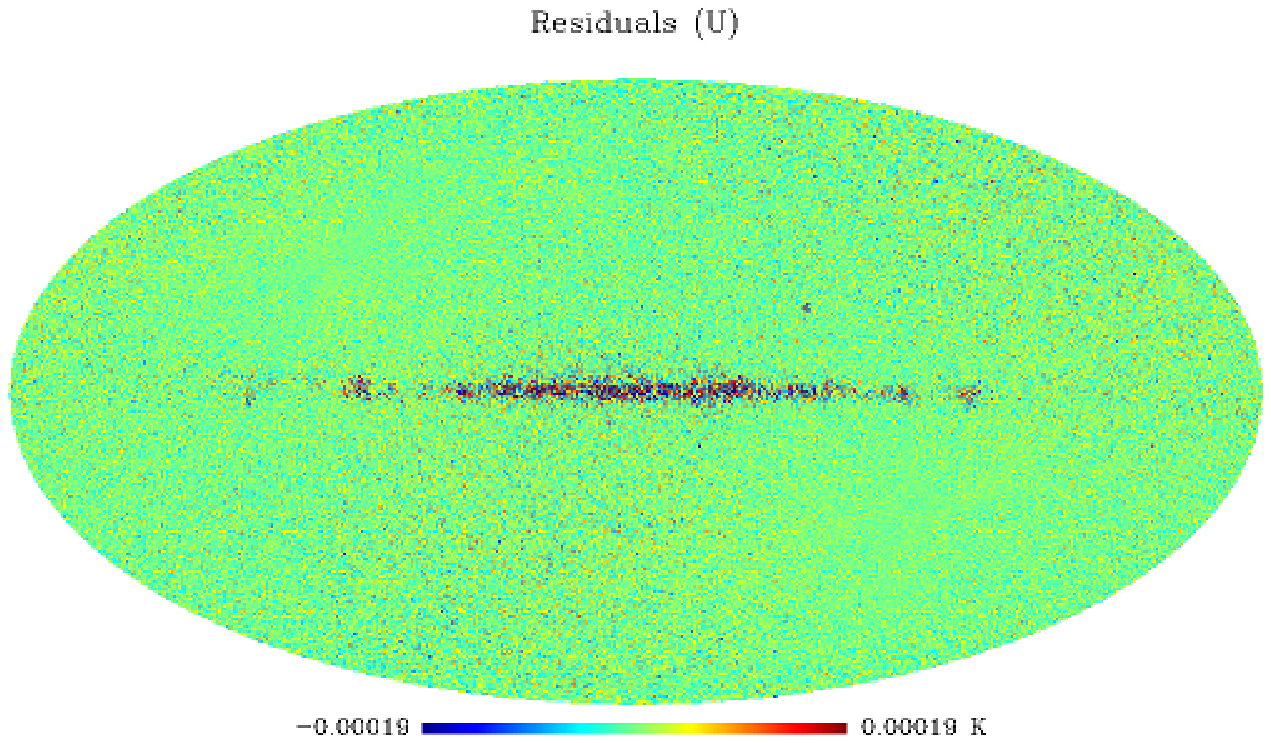}}
%% residuals row
\put(-0.5,-0.5){\includegraphics{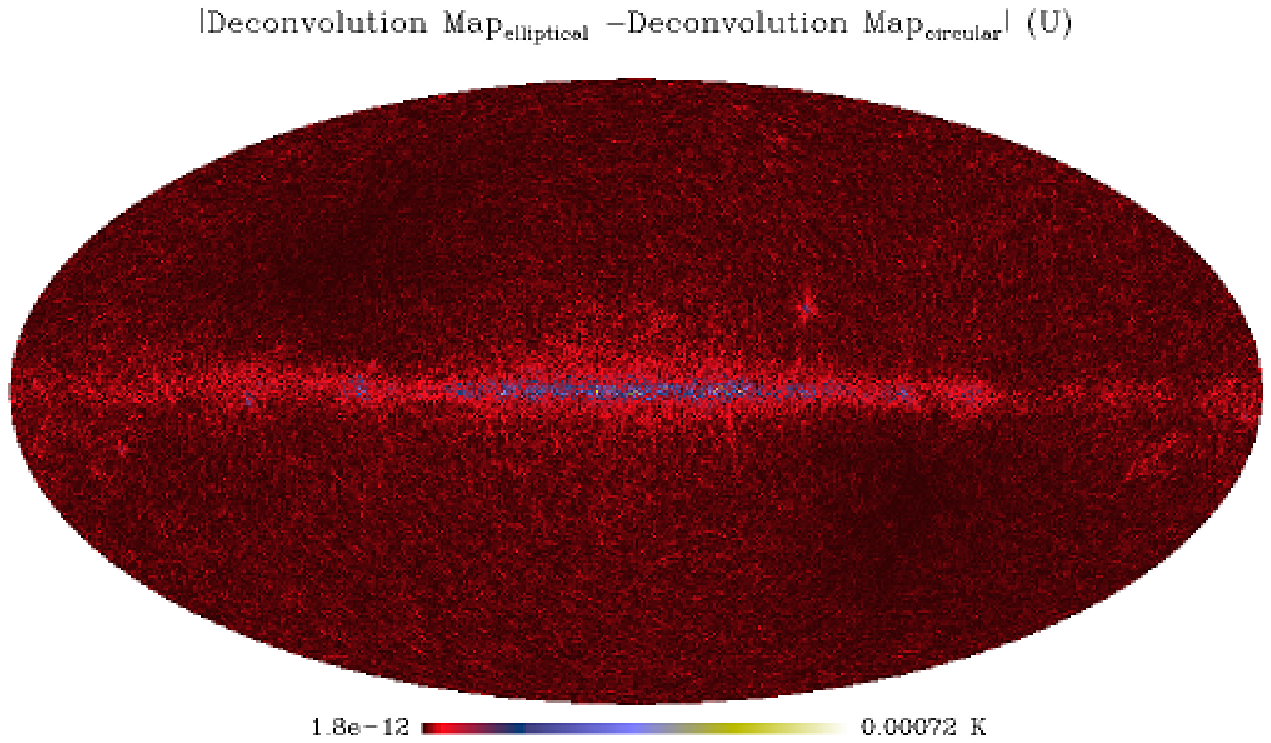}}
\end{picture}
\end{center}
\caption[]{As Figure~\ref{compQresiduals_fig} but for the Stokes U parameter.}
\label{compUresiduals_fig}
\end{figure}
%%%%%%%%%%%%%%%%%%%%%%%%%%%%%%%%%%%%%%%%%%%%%%%%%%%%%%%%%%%%%%%%%%%%%%%%%%%
%%%%%%%%%%%%%%%%%%%%%%%%%%%%%%%%%%%%%%%%%%%%%%%%%%%%%%%%%%%%%%%%%%%%%%%%%%%

%%%%%%%%%%%%%%%%%%%%%%%%%%%%%%%%%%%%%%%%%%%%%%%%%%%%%%%%%%%%%%%%%%%%%%%%%%%
%% Figure with maps for the case of no Pt & SZ sources 
%% but recovered with the symmetric beams-- T
%%%%%%%%%%%%%%%%%%%%%%%%%%%%%%%%%%%%%%%%%%%%%%%%%%%%%%%%%%%%%%%%%%%%%%%%%%%
\begin{figure}
\begin{center}
\setlength{\unitlength}{1cm}
\begin{picture}(8,11)(0,0)
%% map row
\put(-0.5,4.75){\includegraphics{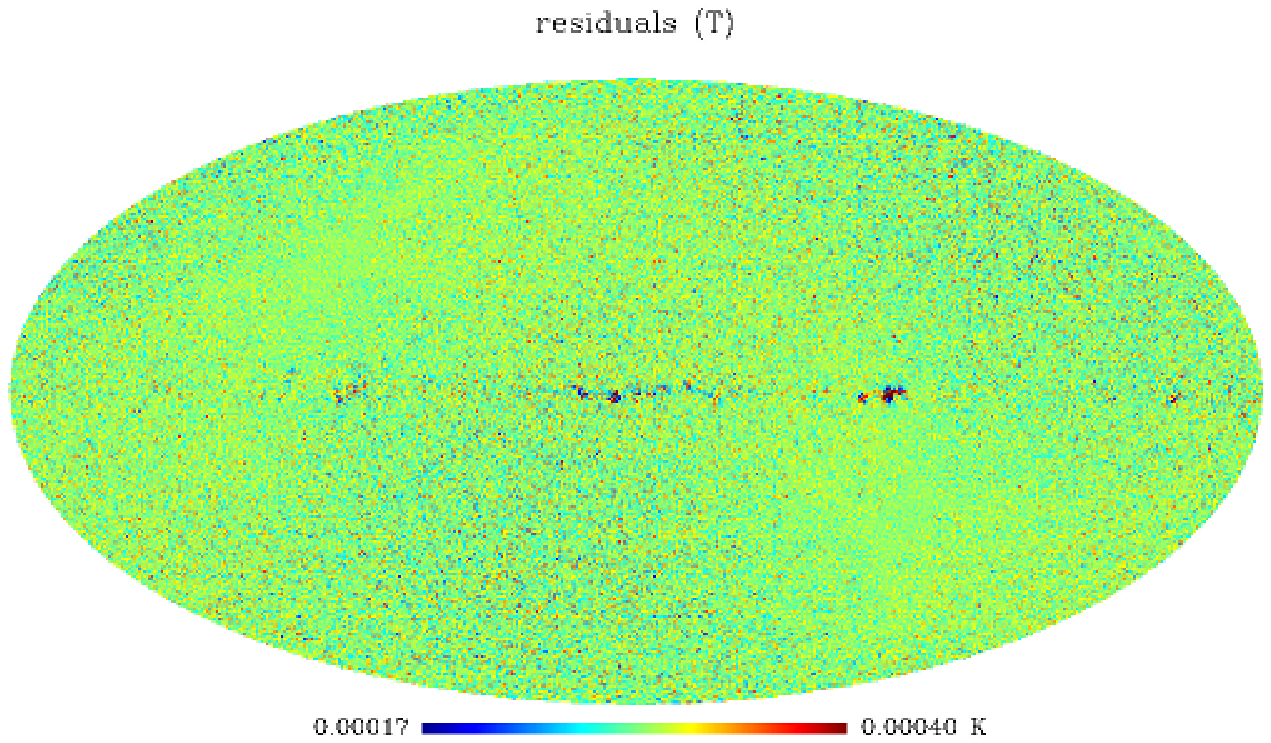}}
%% residuals row
\put(-0.5,-0.5){\includegraphics{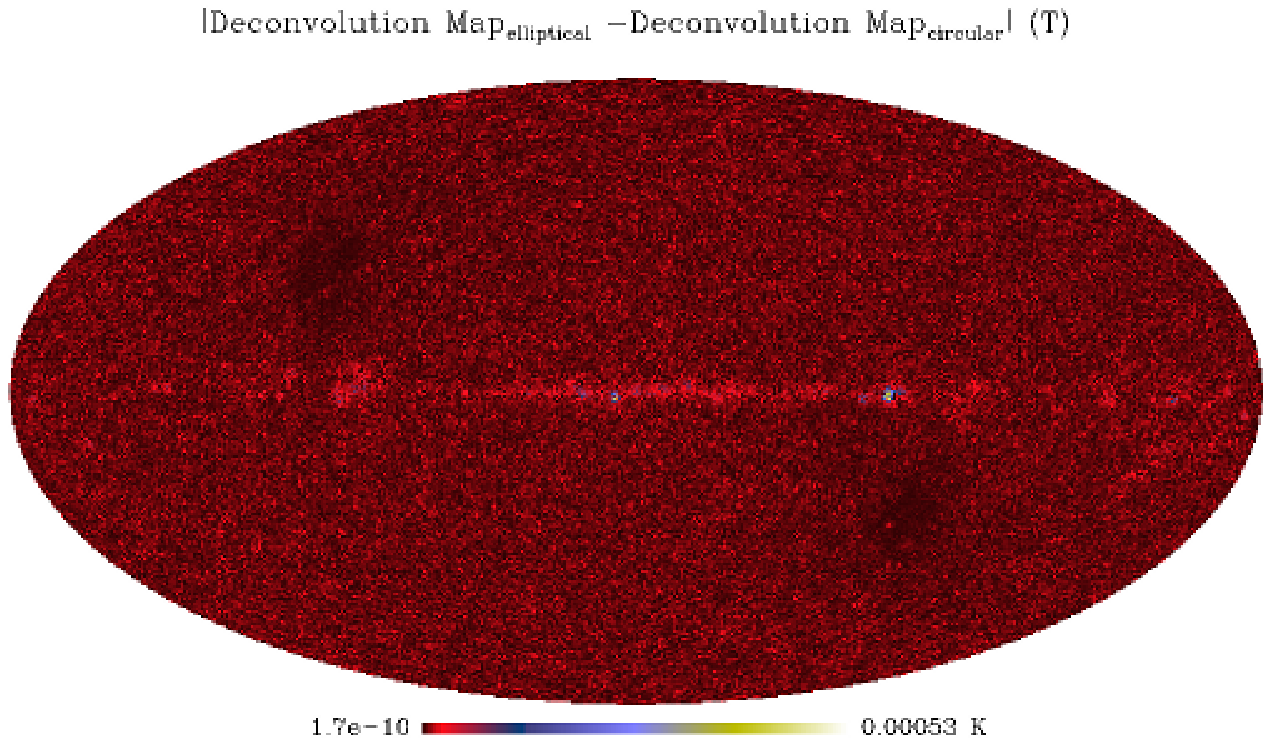}}
\end{picture}
\end{center}
\caption[]{Top: Residuals between the input-sky and the recovered $a_{\ell
    m}^T$, plotted over the same range as the residuals, in
    Figure~\ref{noptsT_fig}. The input-sky, containing only the diffuse
    foregrounds, is the same as that shown in Figure~\ref{noptsT_fig} whereas
    the $a_{\ell m}^T$ were recovered using circular beams rather than the
    elliptical beams used in the simulations. Bottom: Absolute difference between the maps recovered using elliptical and circular beams.}
\label{compTresiduals_nopts_fig}
\end{figure}
%%%%%%%%%%%%%%%%%%%%%%%%%%%%%%%%%%%%%%%%%%%%%%%%%%%%%%%%%%%%%%%%%%%%%%%%%%%

%%%%%%%%%%%%%%%%%%%%%%%%%%%%%%%%%%%%%%%%%%%%%%%%%%%%%%%%%%%%%%%%%%%%%%%%%%%
%% cl figure
%%%%%%%%%%%%%%%%%%%%%%%%%%%%%%%%%%%%%%%%%%%%%%%%%%%%%%%%%%%%%%%%%%%%%%%%%%%
\begin{figure}
\begin{center}
\setlength{\unitlength}{1cm}
\begin{picture}(8,19)(0,0)
\put(9,11.25){\includegraphics{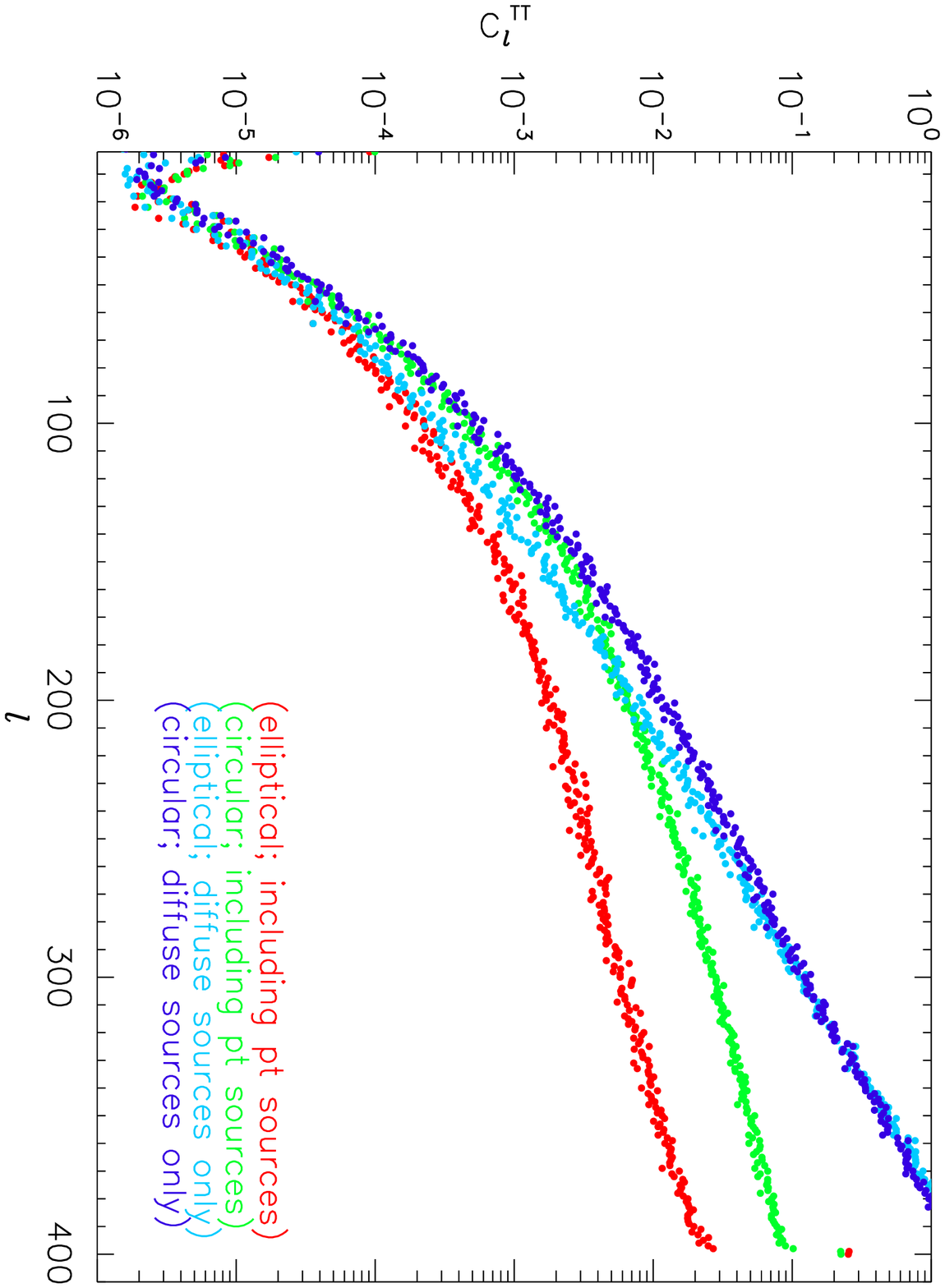}}
\put(9,5){\includegraphics{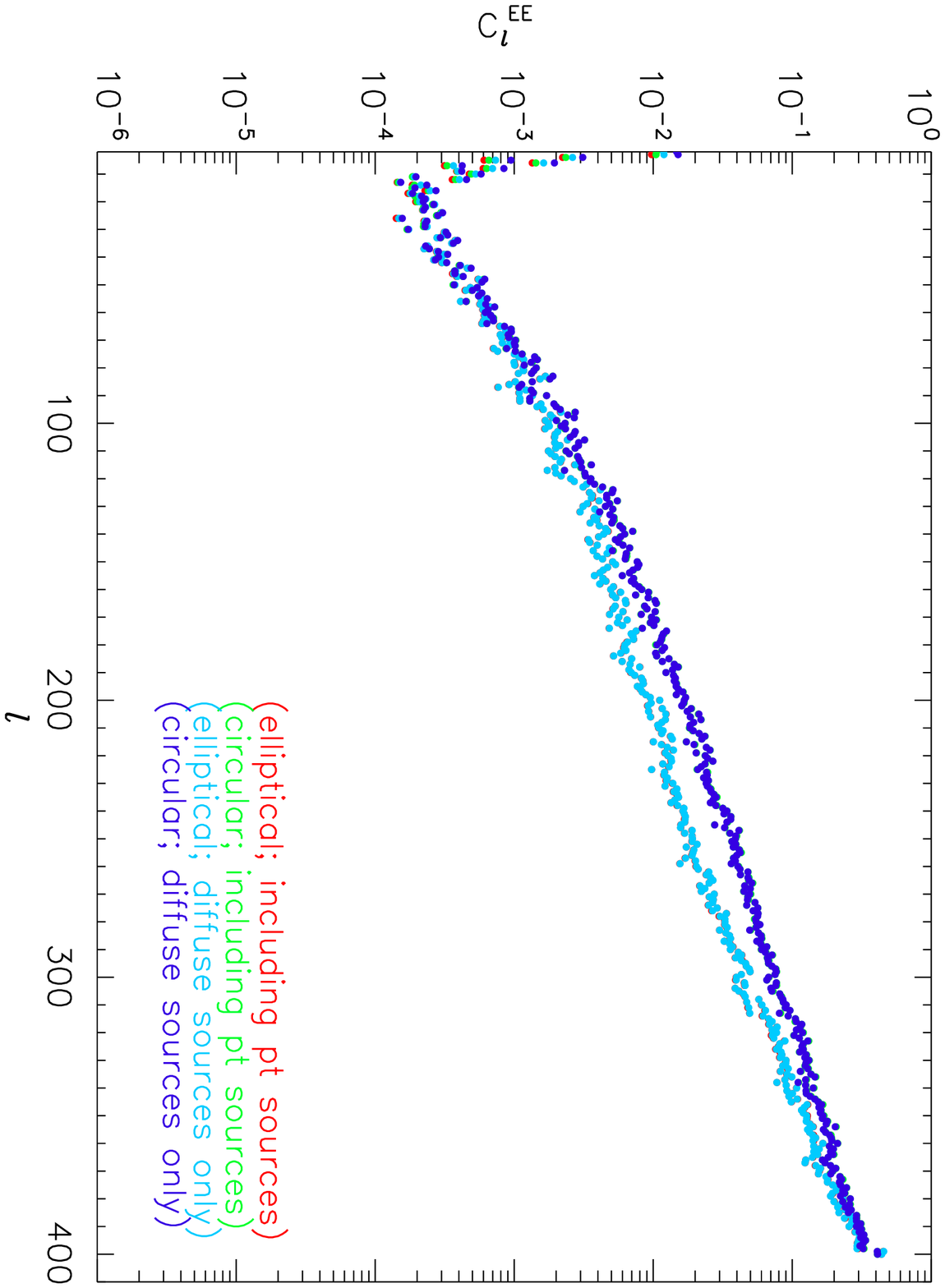}}
\put(9,-1.25){\includegraphics{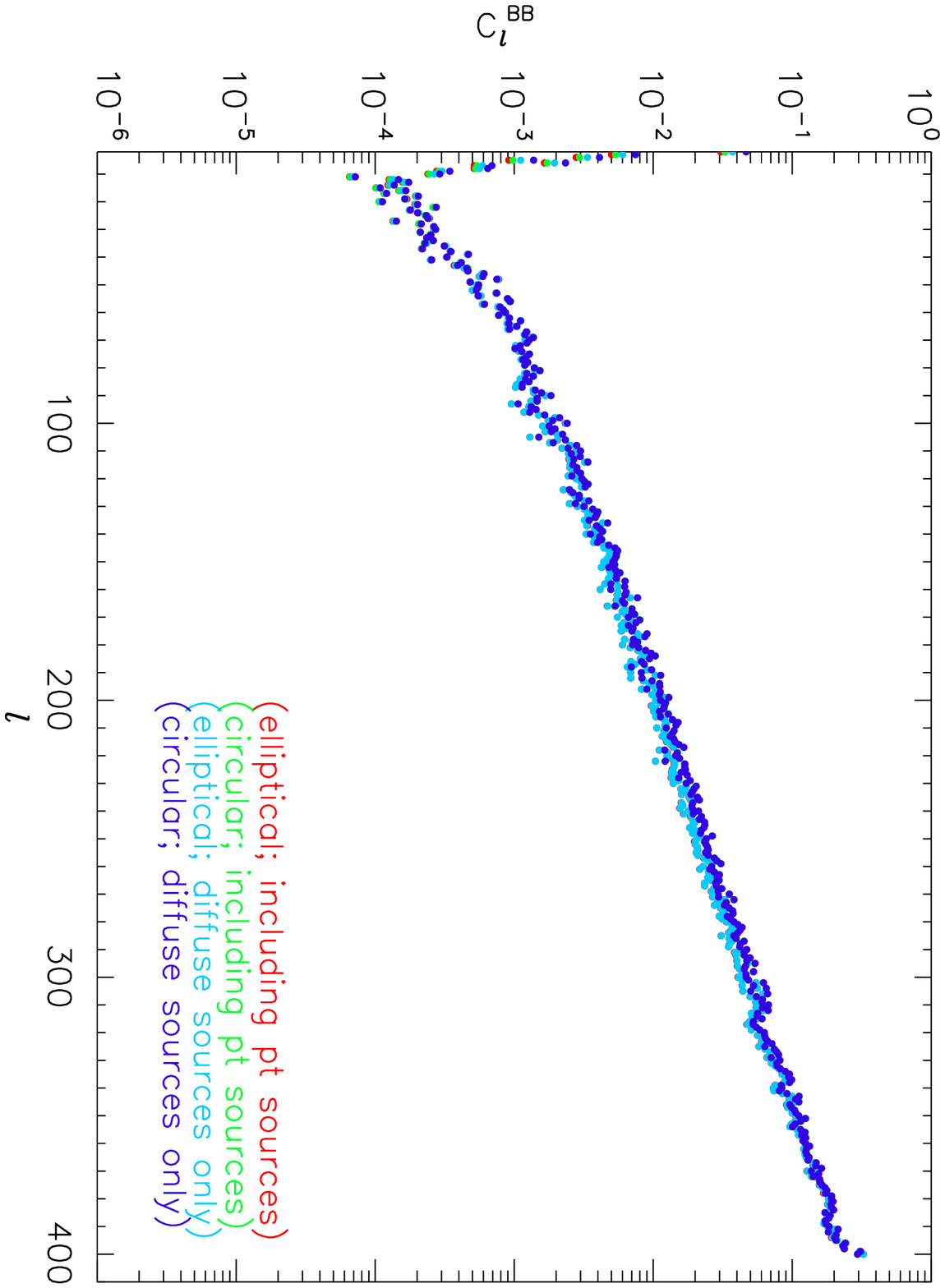}}
\end{picture}
\end{center}
\caption[]{Power spectra of the residuals in $T$ (top), $E$, (middle) and $B$ (bottom) divided by the power spectrum of the input sky. Shown here are the results of the analysis of two different sets of simulated data with two different beams. Both data set were simulated using the elliptical beams, and contain the CMB, diffuse Galactic foregrounds and $1/f$ and white noise. The second simulated data set additionally includes point and SZ sources. Each data set was analysed twice, once using the elliptical beams and once with the circular beams. The dark and light blue points are from the analysis of the first data set, with the circular and elliptical beams respectively. The green and red points correspond to the analysis of the second data set again with the circular and elliptical beams respectively. The differences between the recoveries due to the beams may easily be seen in the figures. In the temperature power spectra the differences are only apparent at intermediate $\ell$-values, for the case where there are no point sources, as at higher $\ell$-values the noise dominates. The up-turn in the fractional residuals in all the figures at low $\ell$ is due to residual $1/f$ noise, that was not removed by discarding the zero-frequency Fourier coefficients from the analysis.}
\label{cls_fig}
\end{figure}
%%%%%%%%%%%%%%%%%%%%%%%%%%%%%%%%%%%%%%%%%%%%%%%%%%%%%%%%%%%%%%%%%%%%%%%%%%%

\section{{\it Trieste} simulations}
\label{trieste_sim}

As well as testing our method on the complex-beam simulations, as described in Section~\ref{own_sim},  we also used a set of simulations produced by the {\it Planck} CTP Working Group, which is concerned with the evaluation of map-making, power-spectrum and likelihood methods for {\it Planck} data. The CTP simulations were produced to enable the comparison of a number of map-making algorithms, \citep{trieste_paper} and they will be referred to throughout this paper as the {\it Trieste} simulations. They were generated using a simulations pipeline developed by the {\it Planck} collaboration with the purpose of providing simulated {\it Planck} data, \citep{levelS_paper} .

One year of pointing and signal data for the LFI 30~GHz channel, which corresponds to 8784 rings of TOD, was generated.  Each repositioning of the spin axis, corresponds to one hour of data and to one ring of TOD, and this repositioning follows a cycloidal scanning strategy. The data were simulated for the two polarised detector pairs of which the LFI 30~GHz channel is comprised. Two different sets of beams, circular and elliptical, were used with these simulations allowing the effects of different
beams to be investigated. The circular beams are Gaussian beams with a full-width half-maximum (FWHM) of $32.5\arcmin$, which is the geometric mean of the two FWHMs of the elliptical beams. These elliptical beams are the best-fit elliptical approximation to the pre-launch LFI~30~GHz beams \citep{sandri10}, for each of the two detectors corresponding to each of the two horns.

These data include the effects of variable spin velocity and nutation. There is also the option of including sampling effects, where the effects of the finite sampling period of the detectors are taken into account.  Here we use the most realistic data, which is, in this case, the TOD simulated using the elliptical beams, in which the effects of sampling are included. The following components of the signal are included in the simulations: the CMB, the diffuse Galactic foregrounds, including the synchrotron, free-free and dust, compact objects, such as Sunyaev-Zel'dovich (SZ) clusters and point sources, and both white noise and $1/f$ noise. The models and templates used for these foregrounds are described in \cite{levelS_paper} and references therein; it should be noted that the templates used for the diffuse Galactic emission are extrapolations to {\it Planck} frequencies.

\subsection{Analysis of {\it Trieste} data}
\label{trieste_results}

In this section we present the results of the analysis, using the  iterative approach of Section~\ref{approx_section}, of  the most realistic sub-set of {\it
 Trieste} simulations; which were created using the cycloidal scanning
 strategy, the elliptical beams and included the effects of sampling. The
 8784 simulated pointing periods, corresponding to one year of data were
 processed and the sky multipoles were reconstructed up to
 $\ell_{max}=400$. It should be noted that the simulations contain signals at
 $\ell$ values greater than $\ell_{max}$. Given the beam sizes of the LFI~30~GHz detectors, and hence their much reduced sensitivity to power above $\ell=400$, this value for $\ell_{max}$ should be sufficient at least in the case of diffuse signals. Indeed, curtailing the reconstruction acts like a low-pass filter preventing the high-frequency noise from dominating the maps, due to the deconvolution.

The first set of simulated data to be analysed included contributions from the
CMB, diffuse Galactic foregrounds, $1/f$ and white noise. The results of the
analysis of these data may be seen in Figures~\ref{noptsT_fig},
\ref{noptsQ_fig}~and~\ref{noptsU_fig}. In these figures the $a_{\ell m}$
corresponding to the signals input to the simulations were used to generate a
map of the input-sky for comparison with the map produced from the recovered
$a_{\ell m}$. In order to perform this comparison the input $a_{\ell m}$ were
curtailed to the same value of $\ell_{max}$ that was used in the recovery of
the  $a_{\ell m}$ from the simulated data. The differences between these two
sets of $a_{\ell m}$ were used to generate residual maps, in order to
illustrate the differences between the recovered and the input sky. In
Figures~\ref{noptsT_fig}~--~\ref{noptsU_fig} it is seen that there has been a
successful reconstruction of the input sky, with the only features visible in
the residual maps being due to the noise modulated by the hit-count. The
hit-count is determined by the scanning strategy and this is the
cause of the lower level of the residuals seen at the ecliptic poles, where the coverage is very much enhanced over that of the rest of the sky. These residuals maps also demonstrate that the remaining unrecovered power at higher multipoles has not interfered with the recovery of the lower multipoles.

The second set of data was the same as the first except that the contributions from SZ and point sources were also included. These simulations were first analysed using the elliptical beams they were produced with and then they were reanalysed assuming circular beams. The recovery of the $a_{\ell m}$ when using
the elliptical beams results in residuals maps indistinguishable from the
residual maps for $Q$ and $U$ from the analysis of the first data set (shown in
Figures~\ref{noptsQ_fig}~and~\ref{noptsU_fig}). The residuals for the $T$ map,
however, show that the recovery is not ideal in the region of bright point
sources. The top panel of Figure~\ref{comp_Tresiduals_fig} shows these
residuals plotted over the same colour range as the residual map for $T$ from
the analysis of the first data set. This sensitivity to point-like objects is
due to the resultant extra power at small scales, and the fact the recovery is limited in $\ell$. The contribution of the point sources to the polarisation signal is small, so the $Q$ and $U$ maps remain unaffected. Increasing the value of $\ell_{max}$ used in the recovery of the $a_{\ell m}$ will remove these effects from the residual map. This sensitivity was not seen in the complex-beam simulations, with their relatively higher value of $l_{max}$ in comparison to the beam size.

The residuals from the analysis of this simulated data set with the circular beams, equivalent to a standard pixel-based map-making analysis, are shown in Figures~\ref{comp_Tresiduals_fig}, \ref{compQresiduals_fig}~and~\ref{compUresiduals_fig}. These residual maps are again plotted over the same colour range as the residual maps produced from the analysis of the first data set, shown in Figures~\ref{noptsT_fig}~--~\ref{noptsU_fig}. Recalling that the residual maps produced from the recovery using the elliptical beams were indistinguishable from the residual maps produced from the analysis of the first data set for the $Q$ and $U$ maps, it is seen that for all the maps there is an increase in the magnitude of the residuals along the Galactic plane and in the vicinity of beam-sized or smaller objects when circular beams are assumed. Figure~\ref{compTresiduals_nopts_fig} shows the residual T map found from the analysis of the first data set with the circular beams; this also shows an increase in the level of the residuals.

The bottom panels of Figures~\ref{comp_Tresiduals_fig}, \ref{compQresiduals_fig},  \ref{compUresiduals_fig}~and~\ref{compTresiduals_nopts_fig} show maps of the absolute difference between the maps recovered using the elliptical beams and the maps recovered using the circular beams. Given that the same set of simulated data is used in each case, all the differences must be due to the difference between the elliptical and circular beams. The smallest differences between these recovered maps are observed to be at the ecliptic poles. This is due to the properties of the sky coverage in these regions, with multiple intersecting scans at many different orientations, and the fact that the circular beams used have FWHMs which are the geometric mean of the two FWHMs which describe each elliptical beam.

It is noticeable that the magnitude of the additional errors in the
recovery due to using the circular, and in this case incorrect, beams is
typically larger than the residuals due to the noise in the data. The differences
between the recoveries due to the different beams may easily be seen in
Figure~\ref{cls_fig} which shows the fractional errors in the power spectrum
of both data sets with both the elliptical and circular beams. This figure
shows the ratio of the power spectra of the residual $a_{\ell m}$, which are
formed from the difference between the input and recovered $a_{\ell m}$, and
the input power spectra. These power spectra are seen to increase with
increasing $\ell$. This increase is observed as the beams have been
deconvolved and is due to the fact that the signal is suppressed by the beams
whereas the noise is not. The dark blue (green) and light blue (red) points
are from the analysis of the first (second) data set, with the circular and
elliptical beams respectively. The first data set does not contain point or SZ
sources whereas the second data set does, the input power spectra for the
second data set will therefore have more power at smaller scales, especially
in $T$, than that of the first data set. This leads to the smaller fractional
errors seen for the analysis of the second data set, at higher multipoles in
the $T$ power spectra. For the $E$ and $B$ power spectra the figure shows virtually no differences between the two different data sets, for each beam. The $B$ power spectra for the different beams are very similar, whereas there are clear differences between the $E$ power spectra formed using the different beams, as may be expected from \cite{carretti04}, who showed that the $E$ power spectrum is coupled to the unpolarised sky through the beam, with the contamination peaking on the scale corresponding to the FWHM of the beam. In the case of the temperature power spectra the differences are only apparent at intermediate $\ell$-values, for the case where there are no point sources, as at higher $\ell$-values the noise dominates. Whereas the differences between the the power spectra formed from the different beams using the second data set, which contained point sources, are visible for all $\ell$ values higher than $\ell \sim 70$.

%%%%%%%%%%%%%%%%%%%%%%%%%%%%%%%%%%%%%%%%%%%%%%%%%%%%%%%%%%%%%%%%%%%%%%%%%%%

\section{Conclusions}
\label{section_conclusions}

We have described a successful implementation of the map-making methodology developed in \cite{challinor02} and \cite{leeuwen02}. While the computational costs of our implementation are far from trivial, the ability to deconvolve any arbitrarily complex beam from the data may prove to be worth the computational expense, at least for an analysis of the {\it Planck} LFI channels. Due to their position at the edges of the focal plane the LFI detectors are anticipated to have more elongated beams than their HFI counterparts.

We have demonstrated that this method may successfully be applied in the presence of compact and point sources provided that recovery proceeds to a value of $\ell_{max}$ at which there is little remaining unresolved signal. We have also shown how ignoring beam asymmetries affects the recovery of the {\it Planck} maps, both for temperature and polarisation. These effects are most noticeable in the Galactic plane region and near the locations of compact or point source objects.  

We have shown that our method can produce maps without the distorting influence of the beams. This property may be especially useful for producing the best possible maps for Galactic science where we have shown that ignoring beam asymmetries will results in larger distortions. 

As the attractiveness of this method increases with increasing beam complexities, it may also be of use for proposed future experiments, which are likely to consist of many more detectors, hence there will be detectors further from the centre of the focal plane resulting in the beams being more distorted. Additionally, the analysis of data from any future experiment would benefit from increases in computing performance, likely to occur in the interim.

%%%%%%%%%%%%%%%%%%%%%%%%%%%%%%%%%%%%%%%%%%%%%%%%%%%%%%%%%%%%%%%%%%%%%%%%%%%%

\begin{acknowledgements}

This work was supported by STFC at the Cambridge Planck Analysis Centre, and utilised COSMOS VI, an Altix 3700 supercomputer, funded by SGI/Intel, HEFCE and PPARC. This research also used resources of the National Energy Research Scientific Computing Center, which is supported by the Office of Science of the U.S. Department of Energy under Contract No. DE-AC02-05CH11231.

Some of the results in this paper have been derived using the HEALPix \citep{gorski05} package. 

This paper has made use of simulated data produced by the CTP Working Group.

\end{acknowledgements}

%%%%%%%%%%%%%%%%%%%%%%%%%%%%%%%%%%%%%%%%%%%%%%%%%%%%%%%%%%%%%%%%%%%%%%%%%%%%%
% BibTeX users please use
\bibliographystyle{aa}
\bibliography{16986}

\begin{thebibliography}{20}
\expandafter\ifx\csname natexlab\endcsname\relax\def\natexlab#1{#1}\fi

\bibitem[{{Armitage} \& {Wandelt}(2004)}]{armitage04}
{Armitage}, C. \& {Wandelt}, B.~D. 2004, \prd, 70, 123007

\bibitem[{{Ashdown} {et~al.}(2007{\natexlab{a}}){Ashdown}, {Baccigalupi},
  {Balbi}, {Bartlett}, {Borrill}, {Cantalupo}, {de Gasperis}, {G{\'o}rski},
  {Heikkil{\"a}}, {Hivon}, {Keih{\"a}nen}, {Kurki-Suonio}, {Lawrence},
  {Natoli}, {Poutanen}, {Prunet}, {Reinecke}, {Stompor}, \&
  {Wandelt}}]{trieste_paper}
{Ashdown}, M.~A.~J., {Baccigalupi}, C., {Balbi}, A., {et~al.}
  2007{\natexlab{a}}, \aap, 471, 361

\bibitem[{{Ashdown} {et~al.}(2007{\natexlab{b}}){Ashdown}, {Baccigalupi},
  {Balbi}, {Bartlett}, {Borrill}, {Cantalupo}, {de Gasperis}, {G{\'o}rski},
  {Hivon}, {Keih{\"a}nen}, {Kurki-Suonio}, {Lawrence}, {Natoli}, {Poutanen},
  {Prunet}, {Reinecke}, {Stompor}, {Wandelt}, \& {The Planck CTP Working
  Group}}]{ctp_mapmaking_two}
{Ashdown}, M.~A.~J., {Baccigalupi}, C., {Balbi}, A., {et~al.}
  2007{\natexlab{b}}, \aap, 467, 761

\bibitem[{{Burigana} {et~al.}(1999){Burigana}, {Malaspina}, {Mandolesi},
  {Danse}, {Maino}, {Bersanelli}, \& {Maltoni}}]{burigana97}
{Burigana}, C., {Malaspina}, M., {Mandolesi}, N., {et~al.} 1999, ArXiv
  Astrophysics e-prints

\bibitem[{{Carretti} {et~al.}(2004){Carretti}, {Cortiglioni}, {Sbarra}, \&
  {Tascone}}]{carretti04}
{Carretti}, E., {Cortiglioni}, S., {Sbarra}, C., \& {Tascone}, R. 2004, \aap,
  420, 437

\bibitem[{{Challinor} {et~al.}(2002){Challinor}, {Mortlock}, {van Leeuwen},
  {Lasenby}, {Hobson}, {Ashdown}, \& {Efstathiou}}]{challinor02}
{Challinor}, A.~D., {Mortlock}, D.~J., {van Leeuwen}, F., {et~al.} 2002,
  \mnras, 331, 994

\bibitem[{{Delabrouille}(1998)}]{delabrouille98}
{Delabrouille}, J. 1998, \aaps, 127, 555

\bibitem[{{Efstathiou}(2005)}]{hybrid_paper}
{Efstathiou}, G. 2005, \mnras, 356, 1549

\bibitem[{{Golub} \& {van Loan}(1996)}]{pcg_ref}
{Golub}, G.~H. \& {van Loan}, C.~F. 1996, {Matrix computations, 3rd edition}
  (London: The Johns Hopkins University Press)

\bibitem[{{G{\'o}rski} {et~al.}(2005){G{\'o}rski}, {Hivon}, {Banday},
  {Wandelt}, {Hansen}, {Reinecke}, \& {Bartelmann}}]{gorski05}
{G{\'o}rski}, K.~M., {Hivon}, E., {Banday}, A.~J., {et~al.} 2005, \apj, 622,
  759

\bibitem[{{Lewis} \& {Challinor}(2006)}]{lensing_review}
{Lewis}, A. \& {Challinor}, A. 2006, \physrep, 429, 1

\bibitem[{{O'Dea} {et~al.}(2007){O'Dea}, {Challinor}, \& {Johnson}}]{odea07}
{O'Dea}, D., {Challinor}, A., \& {Johnson}, B.~R. 2007, \mnras, 292

\bibitem[{{Perotto} {et~al.}(2010){Perotto}, {Bobin}, {Plaszczynski}, {Starck},
  \& {Lavabre}}]{lensing_planck}
{Perotto}, L., {Bobin}, J., {Plaszczynski}, S., {Starck}, J., \& {Lavabre}, A.
  2010, \aap, 519, A4+

\bibitem[{{Planck Collaboration} {et~al.}(2011){Planck Collaboration}, {Ade},
  {Aghanim}, {Arnaud}, {Ashdown}, {Aumont}, {Baccigalupi}, {Baker}, {Balbi},
  {Banday}, \& et~al.}]{Planck_mission11}
{Planck Collaboration}, {Ade}, P.~A.~R., {Aghanim}, N., {et~al.} 2011, ArXiv
  e-prints 1101.2022

\bibitem[{{Poutanen} {et~al.}(2006){Poutanen}, {de Gasperis}, {Hivon},
  {Kurki-Suonio}, {Balbi}, {Borrill}, {Cantalupo}, {Dor{\'e}}, {Keih{\"a}nen},
  {Lawrence}, {Maino}, {Natoli}, {Prunet}, {Stompor}, \&
  {Teyssier}}]{ctp_mapmaking_one}
{Poutanen}, T., {de Gasperis}, G., {Hivon}, E., {et~al.} 2006, \aap, 449, 1311

\bibitem[{{Reinecke} {et~al.}(2006){Reinecke}, {Dolag}, {Hell}, {Bartelmann},
  \& {En{\ss}lin}}]{levelS_paper}
{Reinecke}, M., {Dolag}, K., {Hell}, R., {Bartelmann}, M., \& {En{\ss}lin},
  T.~A. 2006, \aap, 445, 373

\bibitem[{{Sandri} {et~al.}(2010){Sandri}, {Villa}, {Bersanelli}, {Burigana},
  {Butler}, {D'Arcangelo}, {Figini}, {Gregorio}, {Lawrence}, {Maino},
  {Mandolesi}, {Maris}, {Nesti}, {Perrotta}, {Platania}, {Simonetto}, {Sozzi},
  {Tauber}, \& {Valenziano}}]{sandri10}
{Sandri}, M., {Villa}, F., {Bersanelli}, M., {et~al.} 2010, \aap, 520, A7+

\bibitem[{{Tauber} {et~al.}(2010){Tauber}, {Mandolesi}, {Puget}, {Banos},
  {Bersanelli}, {Bouchet}, {Butler}, {Charra}, {Crone}, {Dodsworth}, \&
  et~al.}]{Planck_mission10}
{Tauber}, J.~A., {Mandolesi}, N., {Puget}, J., {et~al.} 2010, \aap, 520, A1+

\bibitem[{{van Leeuwen}(2007)}]{newhip}
{van Leeuwen}, F. 2007, {{H}ipparcos, the {N}ew {R}eduction of the {R}aw
  {D}ata}, {A}strophysics and {S}pace {S}cience {L}ibrary. {V}ol. 350 edn.
  (Springer)

\bibitem[{{van Leeuwen} {et~al.}(2002){van Leeuwen}, {Challinor}, {Mortlock},
  {Ashdown}, {Hobson}, {Lasenby}, {Efstathiou}, {Shellard}, {Munshi}, \&
  {Stolyarov}}]{leeuwen02}
{van Leeuwen}, F., {Challinor}, A.~D., {Mortlock}, D.~J., {et~al.} 2002,
  \mnras, 331, 975

\end{thebibliography}
%%%%%%%%%%%%%%%%%%%%%%%%%%%%%%%%%%%%%%%%%%%%%%%%%%%%%%%%%%%%%%%%%%%%%%%%%%%%%

\begin{appendix}
\section{Implementation Details}
\label{implementation_details}
\subsection{Scaling with number of processors}
This appendix describes the effect, due to the method used to evaluate $R$, on
the scaling with the number of processers used in the parallel implementations
of both the iterative and direct methods. 

\begin{figure}
\begin{center}
\setlength{\unitlength}{1cm}
\begin{picture}(8,11)(0,0)
\put(8,4.5){\includegraphics{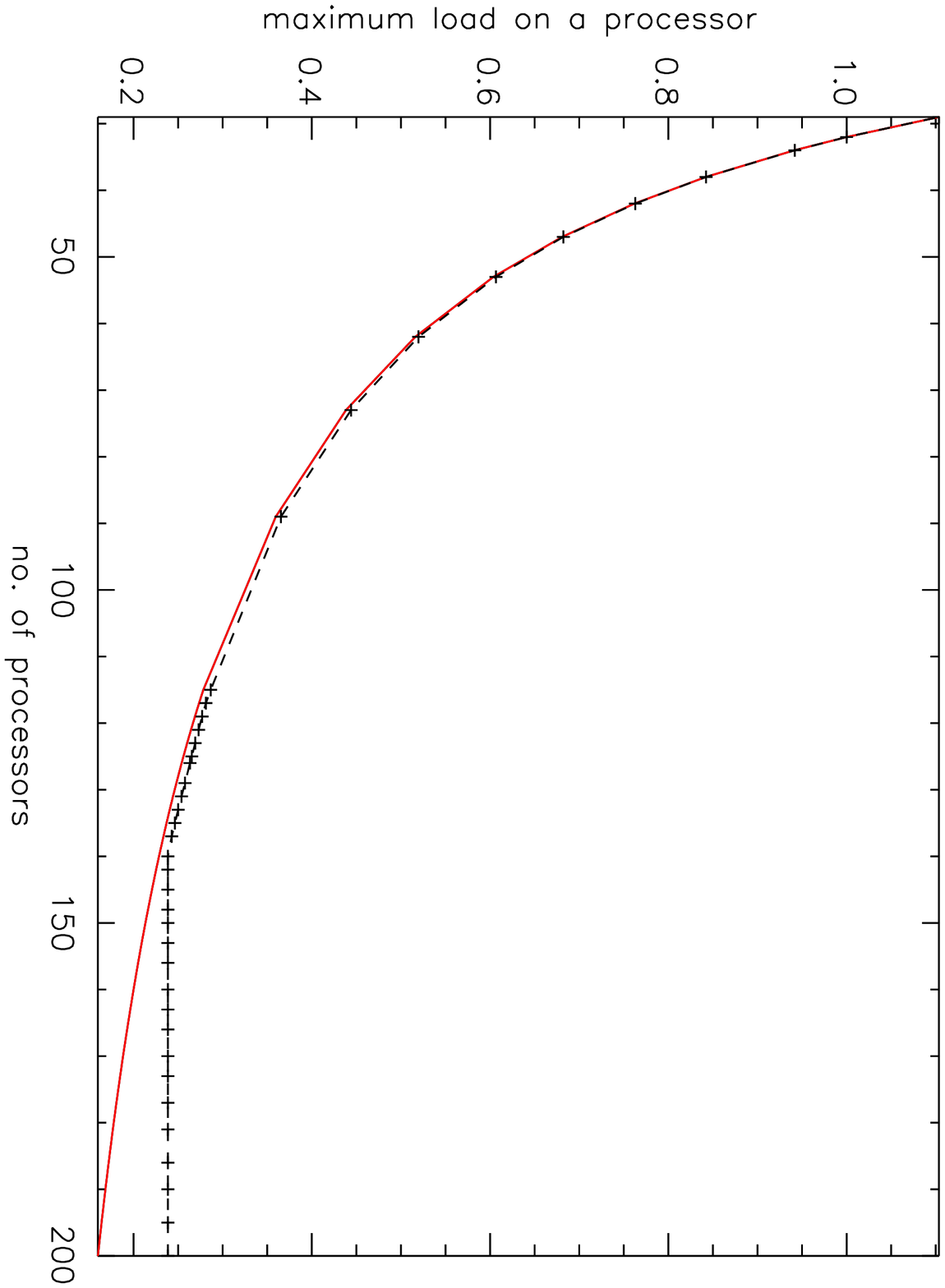}}
\put(8,-0.75){\includegraphics{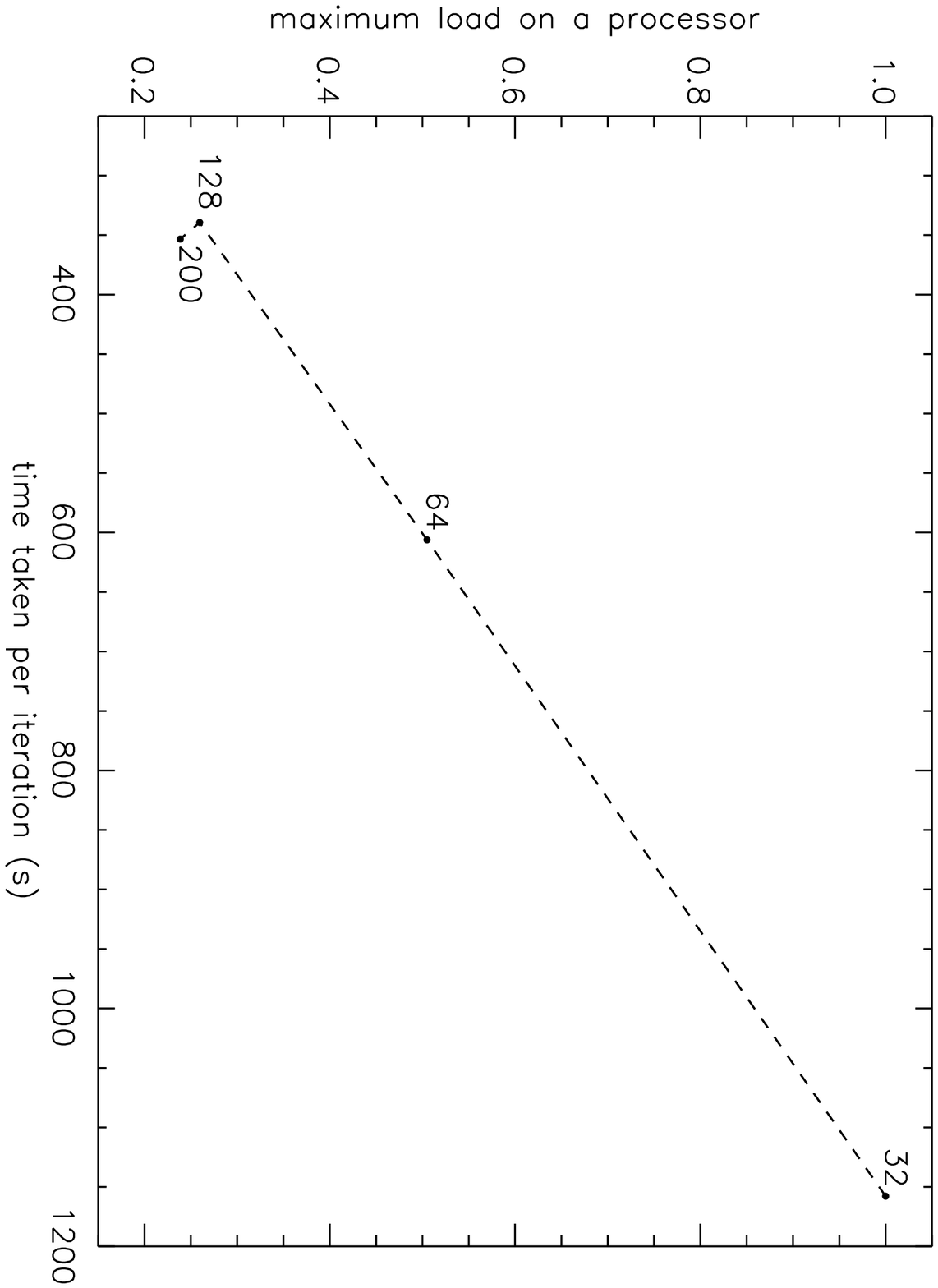}}
\end{picture}
\end{center}
\caption[]{Scaling of the map-making algorithm with the number of processors used. Since each processor is assigned a number of different values of $\ell$, the number of processors to which the code will successfully scale will depend on the value of $\ell_{max}$ to which the analysis is to be performed. For a given $\ell_{max}$ there will be a number of processors above which it is not possible to balance the load between them. This is shown in the top plot for $\ell_{max}=400$. The black crosses show the load on the processor with the heaviest workload, normalised by the equivalent value for 32 processors. The red curve shows the reduction in the workload assuming perfect load balancing. It is seen that there is no reduction in the maximum load once the number of processors is increased beyond 142. The bottom plot shows the relationship between the maximum load and the time required per iteration of the code, where the dashed line shows the expected relationship and the black points, labeled with the number of processor used, show the observed behaviour. }
\label{scale_pcg_fig}
\end{figure}

The time taken per iteration of the code will be determined by the processor with the largest workload. It is therefore important that the load is balanced as equally as possible over the number of processors being used. As the code is parallelised by dividing the individual values of $\ell$ between the various processes, there must come a point at which using additional processors will no longer produce a decrease in the time required, as it becomes impossible to share the workload between the different processors. This situation is shown in the top panel in Figure~\ref{scale_pcg_fig}, for an analysis up to an $\ell_{max}=400$. In this figure the black crosses represent the load on the processor with the maximum load, relative to the processor with the maximum load in the case when 32 processors are used. The red curve shows, for comparison, the reduction in the workload, with increasing numbers of processors, assuming perfect load balancing. The bottom panel in Figure~\ref{scale_pcg_fig} shows the relationship between the maximum load and the time taken per iteration. Doubling the number of processors used halves the workload per processor which in turn halves the time taken per iteration, up until the point at which the load balancing breaks down. The number of processors at which this occurs may be easily evaluated and is found to be equal to $\ell_{max}/2.8$.
%%
%%%%%%%%%%%%%%%%%%%%%%%%%%%%%%%%%%%%%%%%%%%%%%%%%%%%%%%%%%%%%%%%%%%%%%%%%%%%%
%%%%%%%%%%%%%%%%%%%%%%%%%%%%%%%%%%%%%%%%%%%%%%%%%%%%%%%%%%%%%%%%%%%%%%%%%%%%
\end{appendix}

%%%%%%%%%%%%%%%%%%%%%%%%%%%%%%%%%%%%%%%%%%%%%%%%%%%%%%%%%%%%%%%%%%%%%%%%%%%%%
\end{document}